\documentclass[10pt,journal,compsoc]{IEEEtran}
\usepackage{url}
\usepackage{multirow}
\usepackage{colortbl}
\usepackage[table,xcdraw]{xcolor}
\usepackage{cite}
\usepackage{amsmath,amssymb,amsfonts}
\usepackage{algorithmic}
\usepackage{graphicx}
\usepackage{textcomp}
\usepackage{booktabs}
\usepackage{mathrsfs}

\begin{document}
%
\title{Improving Representation of High-frequency Components for Medical Visual Foundation Models}

\author{Yuetan Chu, Yilan Zhang, Zhongyi Han, Changchun Yang, Longxi Zhou, Gongning Luo, \textit{Member, IEEE} Chao Huang, Xin Gao, \textit{Senior Member, IEEE}
\thanks{Corresponding authors: Xin Gao; Gongning Luo; Chao Huang}
\thanks{This publication is based upon work supported by the King Abdullah University of Science and Technology (KAUST) Office of Research Administration (ORA) under Award No REI/1/5234-01-01, REI/1/5414-01-01, REI/1/5289-01-01, REI/1/5404-01-01, REI/1/5992-01-01, URF/1/4663-01-01, Center of Excellence for Smart Health (KCSH), under award number 5932, and Center of Excellence on Generative AI, under award number 5940.}
\thanks{Yuetan Chu and Yilan Zhang contributed equally to this paper.}
\thanks{Yuetan Chu, Yilan Zhang, Zhongyi Han, Changchun Yang, Longxi Zhou, Gongning Luo, Xin Gao are with the Computer Science Program, Computer, Electrical and Mathematical Sciences and Engineering Division; Center of Excellence for Smart Health (KCSH), and Center of Excellence on Generative AI, King Abdullah University of Science and Technology (KAUST), Thuwal, Saudi Arabia. (email: xin.gao@kaust.edu.sa, gongning.luo@kaust.edu.sa).}
\thanks{Chao Huang is with the Ningbo Institute of Information Technology Application, Chinese Academy of Sciences (CAS), Ningbo, China. (email: chuang@ict.ac.cn).}}

%



\IEEEtitleabstractindextext{%
\begin{abstract}
Foundation models have recently attracted significant attention for their impressive generalizability across diverse downstream tasks. However, these models are demonstrated to exhibit great limitations in representing high-frequency components and fine-grained details. In many medical imaging tasks, the precise representation of such information is crucial due to the inherently intricate anatomical structures, sub-visual features, and complex boundaries involved. Consequently, the limited representation of prevalent foundation models can result in significant performance degradation or even failure in these tasks. To address these challenges, we propose a novel pretraining strategy for both 2D images and 3D volumes, named Frequency-advanced Representation Autoencoder (Frepa). Through high-frequency masking and low-frequency perturbation combined with embedding consistency learning, Frepa encourages the encoder to effectively represent and preserve high-frequency components in the image embeddings. Additionally, we introduce an innovative histogram-equalized image masking strategy, extending the Masked Autoencoder (MAE) approach beyond ViT to other architectures such as Swin Transformer and convolutional networks. We develop Frepa across nine medical modalities and validate it on 32 downstream tasks for both 2D images and 3D volume data. Without fine-tuning, Frepa can outperform other self-supervised pretraining methods and, in some cases, even surpasses task-specific foundation models. This improvement is particularly significant for tasks involving fine-grained details, such as achieving up to a +15$\%$ increase in dice score for retina vessel segmentation and a +8$\%$ increase in IoU for lung tumor detection. Further experiment quantitatively reveals that Frepa enables superior high-frequency representations and preservation in the embeddings, underscoring its potential for developing more generalized and universal medical image foundation models.
\end{abstract}

}

\maketitle

\section{Introduction}
Recently, the rapid development of medical foundation models has attracted considerable attention and emerged as a novel paradigm in artificial intelligence (AI). These models typically involve pretraining on unlabeled datasets using self-supervised learning techniques~\cite{mae, mfm, clip, usfm} or employ paired images, annotations, and text for supervised training~\cite{sam_1, meddr}. By training on large-scale datasets, foundation models can exhibit emergent capabilities~\cite{emergent} and facilitate improved generalization by transferring the knowledge acquired during the pretraining phase to specific tasks or data sources. In the field of healthcare, many foundation models have achieved significant progress across various medical tasks~\cite{usfm, sam_1, sam_2, meddr, xie}, marking a substantial advancement in medical AI research.

\begin{figure}[!t]
\centering
\centerline{\includegraphics[width=0.95\columnwidth]{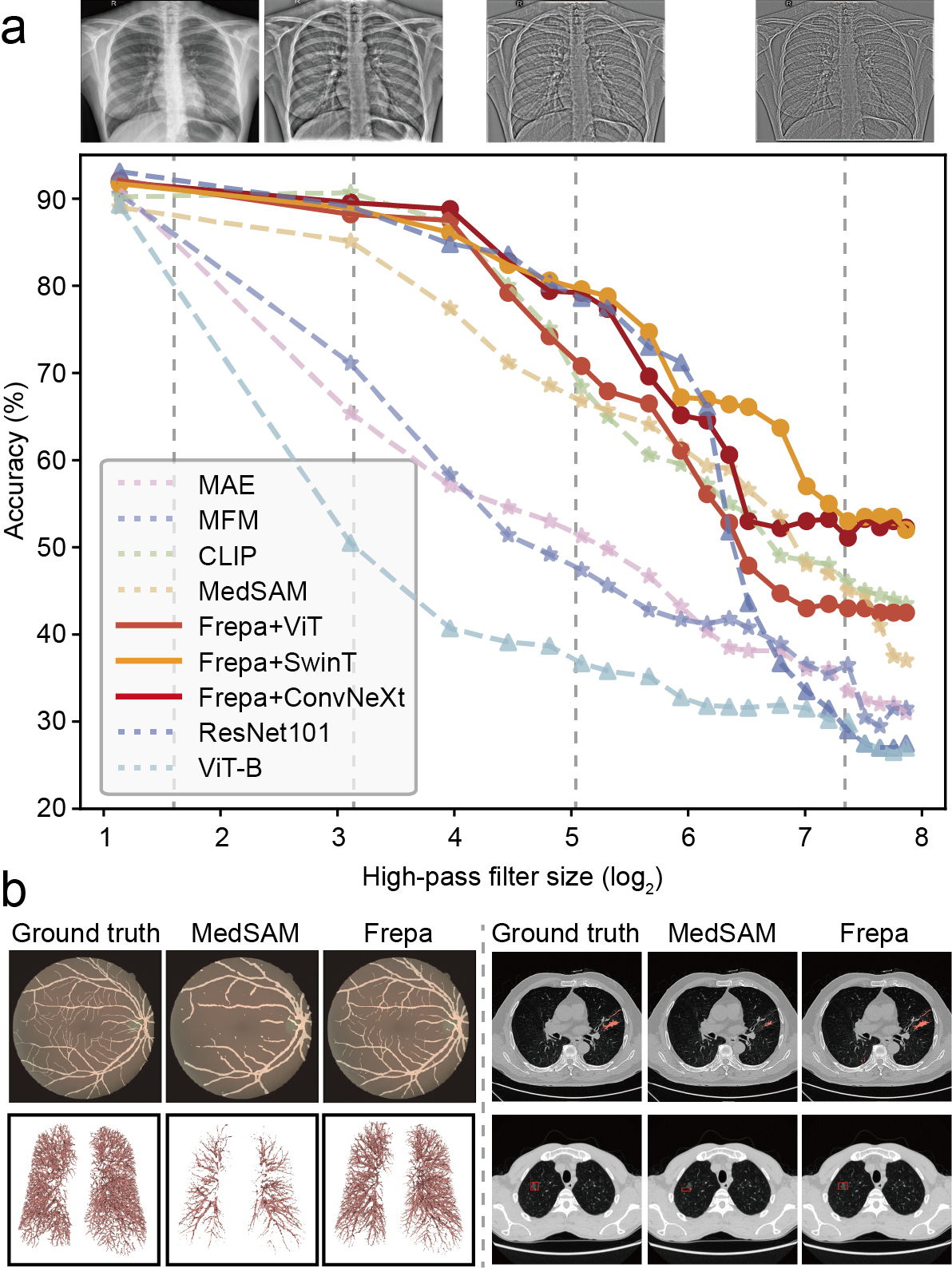}}
\caption{Comparison of the ability of high-frequency representation (a) Classification accuracy on high-pass filtered test set. Initially, models are trained on the raw images. Subsequently, a high-pass filter is applied to the test images to progressively remove low-frequency components, and the classification accuracy of these low-frequency corrupted images is evaluated. Other models exhibit a significant decrease in accuracy as the filtering size increases, indicating limitations in representing high-frequency components, while our method (Frepa) maintains more robust performances. The filtered images are shown at the top. (b) Segmentation results for tasks involving fine-grained details. From top left to bottom right: retinal vessel segmentation, pneumonia segmentation, pulmonary artery segmentation, lung nodule detection. Even after the decoder is re-trained, the segmentation-specific foundation model, MedSAM, still demonstrates poor performance for fine-grained segmentation tasks.}
\label{moti}
\end{figure}

However, both our findings and previous works suggest that these foundation models often result in limited representations of high-frequency components and fine-grained details~\cite{sam_1, scale, de_high_2}. Fig.~\ref{moti} provides two illustrative examples. Fig.~\ref{moti}a shows the classification accuracy of different models on the MedMNIST dataset~\cite{medmnist} after applying high-pass filters of varying sizes. The performance of these models degrades rapidly with increasing filter size, suggesting an over-reliance on low-frequency components and a limited ability to capture high-frequency components. Fig.~\ref{moti}b showcases the segmentation and detection results of tasks involving fine-grained details using the fine-tuned MedSAM~\cite{sam_1} and our proposed methods. This highlights that even segmentation-specific foundation models struggle with fine-grained structures. In diagnostic practices, many diseases and lesions manifest as subtle differences from normal tissue or other disease subtypes, making accurate disease identification heavily reliant on detailed information confined to small regions~\cite{eraly_stage}. Moreover, anatomical structures such as small lesions, blood vessels, and fibrotic tissues, which are critical for both physiological functions and disease diagnosis~\cite{CREMI, av}, also involve high-frequency and fine-grained information. The limited representation of such information is an inherent shortcoming of current foundation models, leading to poor performance or even failure in these tasks and reducing their generalizability to real-world medical applications.

The limitations in high-frequency representation are primarily due to the dominant role of global information in current training strategies. For instance, autoencoder pretraining methods like the masked autoencoder (MAE)~\cite{mae} typically produce reconstructions that capture only general semantic information while lacking specific details~\cite{de_high_2}. In fact, during the reconstruction process, low-frequency, global information tends to be more substantial, causing the model to emphasize these components over high-frequency details~\cite{de_high_1}, resulting in limited detailed information. A similar issue is observed in contrastive learning-based methods~\cite{de_high_4}. Techniques such as CLIP~\cite{clip} rely on textual information and label-level annotations~\cite{meddr}, which contain more global-level information than pixel-level details, leading to a neglect of detailed information capturing. Additionally, as the commonly used backbone network, the vanilla vision transformer (ViT) exacerbates this issue. Compared to convolutional operators (e.g., ConvNeXt~\cite{convnext}) and multi-scale feature maps (e.g., Swin Transformer, SwinT~\cite{swin}), ViT has relatively weaker inductive biases, which results in a reduced ability to capture high-frequency components~\cite{hat, pah}. This is also evident from Fig.~\ref{moti}a. Consequently, the inefficiency in capturing high-frequency information presents significant challenges for current foundation models in medical imaging tasks.


Motivated by the aforementioned observations, we introduce an innovative pretraining strategy for foundational models, which is applicable for both 2D images and 3D volumes, termed the \textbf{Fre}quency-advanced re\textbf{p}resentation \textbf{a}utoencoder (Frepa). Frepa incorporates a novel imaging corruption and reconstruction branch designed to improve the utilization and representation of high-frequency components. This branch, termed frequency dual-component masking, includes two key steps: 1) \textbf{High-frequency domain masking}: We randomly mask high-frequency domains in the frequency spectrum. During training, the image encoders use the remaining information to reconstruct the masked frequency components, encouraging the encoders to learn and leverage realistic high-frequency distributions. 2) \textbf{Low-frequency perturbation}: To prevent the image encoders from relying excessively on easily learned low-frequency components, we introduce a zero-mean perturbation to the low-frequency part of the spectrum, prompting the encoders to rely more heavily on the high-frequency components. Additionally, we propose a hierarchical frequency-to-spatial loss and design an embedding learning strategy to achieve a more robust image reconstruction. We also modify the MAE by incorporating histogram-equalized masking methods, extending its application beyond vanilla ViT to convolutional networks and multi-scale ViT architectures. Furthermore, we propose to integrate the 2D pretrained encoder with iterative intra- and inter-image transformer blocks, enabling it direct application to 3D volume data without retraining. Frepa is developed on 17 million images and validated on 32 downstream tasks across 9 modalities. Frepa not only demonstrates superior performance in general medical tasks but also shows more significant improvement in tasks involving fine-grained information, where other models yield almost failed results. For instance, Frepa outperforms the best other models by approximately 15$\%$ in dice similarity coefficient (DSC) for retina vessel segmentation, 12$\%$ in DSC for pulmonary artery-vein segmentation, and 8$\%$ in Intersection over Union (IoU) for lung tumor detection. Additionally, we also design a quantitative experiment to directly demonstrate that Frepa enables better representation and preservation of high-frequency components in the image embedding compared to other autoencoder pretraining methods. These results showcase that Frepa outperforms in capturing high-frequency components and fine-grained details without compromising the representation of global-level semantic information. Our codes are publicly available as \url{https://github.com/Arturia-Pendragon-Iris/Frepa}.

Our contributions to the state-of-the-art (SOTA) are as follows:
\begin{itemize}
    \item We propose Frepa, a novel approach that significantly enhances the high-frequency capturing capability of foundation models without compromising their low-frequency representation.
    \item We extend the MAE beyond vanilla ViT to improve its generalizability. Additionally, we extend pretrained 2D encoders for direct deployment on 3D volume data.  
    \item We develop Frepa on 17 million images and validated it on 32 downstream tasks for both 2D and 3D data, without requiring fine-tuning.
    \item Extensive experiments demonstrate that Frepa outperforms SOTA methods on most downstream tasks, particularly those involving fine-grained information. Frepa also exhibits exceptional performance and generalizability on previously unseen modalities.
\end{itemize}

\section{Related Work}
\subsection{Foundation model}
The recent advancements in foundation models signify a transformative leap in deep learning, showcasing exceptional generalization and transfer learning capabilities. The multi-modal visual foundation models based on contrastive learning, such as CLIP~\cite{clip}, Flamingo~\cite{flamingo}, and LLaVA~\cite{llava}, have made significant strides in efficient transfer learning and exhibit impressive zero-shot performance. These models, however, rely on large-scale image-text paired datasets and are limited in their applicability to other visual tasks like segmentation or image restoration. On the other hand, autoencoder pretraining strategies enable the network to learn meaningful feature representations from large-scale data in a self-supervised manner via reconstructing raw images from corrupted images. Among these approaches, MAE~\cite{mae} is particularly notable, which pretrains a ViT encoder by randomly masking parts of an image and reconstructing the original image from the remaining patches. However, recent studies have highlighted the limitation of autoencoders in that they tend to focus more on low-frequency information and fail to effectively utilize the full spectrum of frequency information~\cite{scale, de_high_2}. This issue is especially critical in medical imaging tasks, where accurate diagnosis and numerous physiological structures often depend on high-frequency components and fine-grained details. To the best of our knowledge, Frepa is the first study to systematically investigate the frequency representation capabilities of foundation models and to validate the feasibility in the medical field.

\subsection{Medical foundation model}
As foundation models for natural images have achieved significant success, their applications in the medical field are gaining increasing attention. Current research on medical foundation models can be broadly classified into specialist models and generalist models. Specialist models are typically designed to address clinical tasks within a single imaging modality or specific disease type. Examples include USFM~\cite{usfm} for ultrasound, RETFound~\cite{RETFound} for retinal images, SkinGPT-4~\cite{skingpt} for skin disease, M3FM~\cite{chestct} and CT-CLIP~\cite{ctclip} for chest computed tomography, and EndoFM~\cite{endo} for endoscopy videos. These models are typically designed upon the unique characteristics of their respective modalities and are capable of handling diverse downstream tasks. By leveraging knowledge learned during the pretraining phase on large-scale datasets, specialist models often outperform single-modal task-specific models. They can also be flexibly adapted to new tasks using only a small dataset~\cite{chestct}, making them highly efficient for medical applications.

While specialist models often achieve higher accuracy and efficiency within specific modalities, they frequently struggle to generalize across different modalities. In contrast, generalist models demonstrate better generalizability across diverse modalities. Early studies have explored the application of general foundation models, such as GPT-4~\cite{founda_3}, for clinical findings and diagnosis. Subsequently, medical generalist models designed for computer-aided diagnosis, such as BiomedGPT~\cite{BiomedGPT}, MedDr~\cite{meddr}, Medical Gemini~\cite{gemini}, and LLAVA~\cite{llava}, were developed using public medical datasets to scale and integrate clinical knowledge. These models have shown robust capabilities in disease diagnosis, question answering, and medical report generation. Such approaches typically rely on large-scale training data, extensive computational resources, and numerous parameters, driving advancements in generalization. Additionally, generalist models targeting other tasks have also achieved notable success, including MedSAM~\cite{sam_1} for semantic segmentation and MINIM~\cite{MINIM} for image generation. However, in the process of transferring pretraining techniques, these models often do not adequately consider the data characteristics of medical images, such as fine anatomical structures, multi-scale objects, and complex, uncertain boundaries. Consequently, they may perform poorly on medical imaging tasks that depend on fine-grained details and high-frequency information ~\cite{sam_1, meddr} (Fig.~\ref{moti}), posing a significant barrier to their practical application in clinical settings. In this work, we focus on lightweight generalist models and introduce a novel pretraining strategy designed to enhance the ability of foundation models to represent high-frequency information. Our approach aims to improve performance on fine-grained tasks, thereby extending the practical applicability and generalizability of these models in clinical practice.

\section{Methods}
\begin{figure*}[!t]
\centerline{\includegraphics[width=\textwidth]{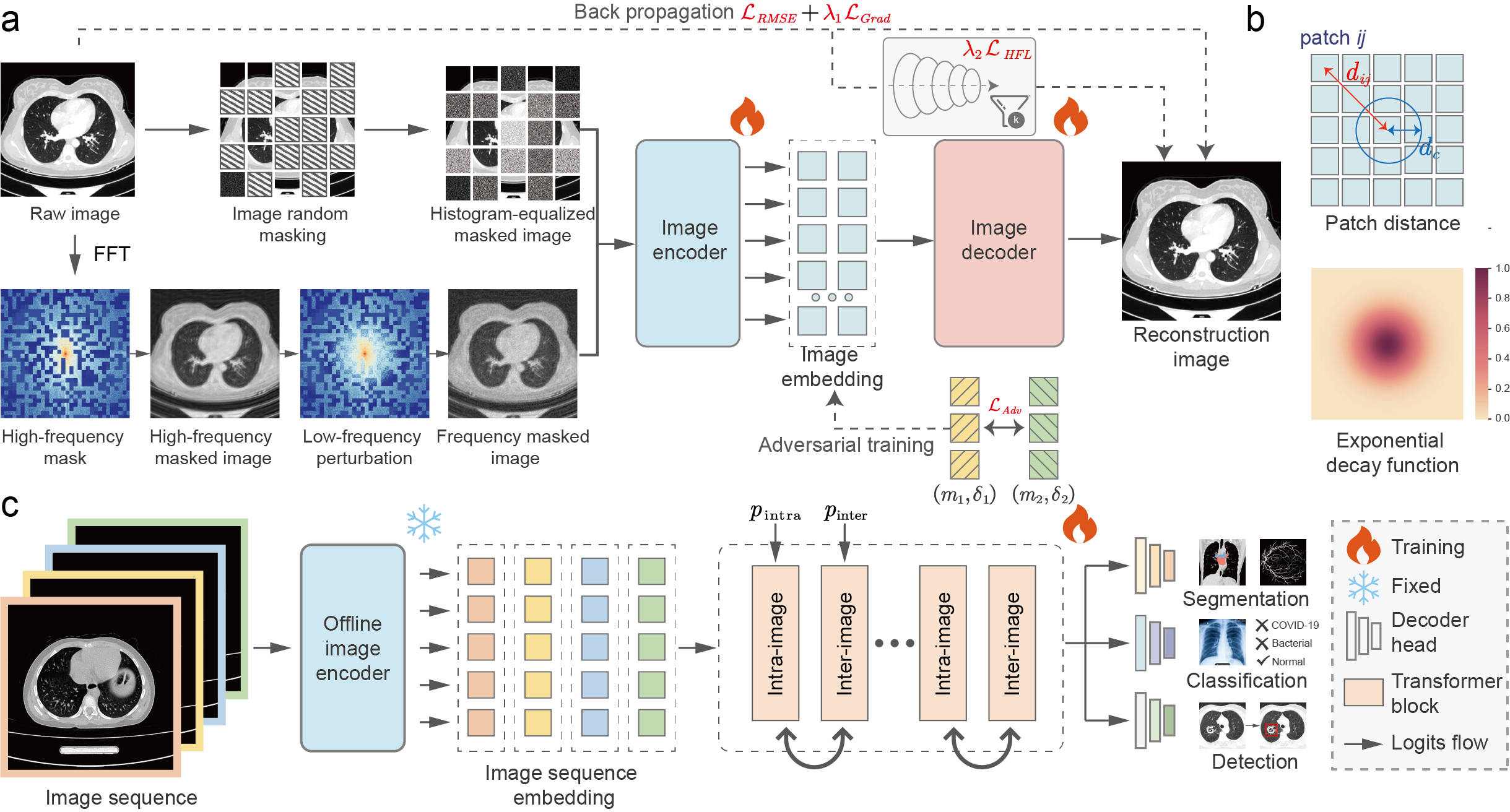}}
\caption{Overall architecture of the proposed Frepa. (a) The pretrained Frepa employs two parallel strategies: frequency dual-component masking and histogram-equalized masking. These strategies are applied to the original image with an equal probability to obtain the corrupted image. (b) Illustrations of the distance calculation and the exponential decay function used in the frequency dual-component masking. (c) The method for extending the 2D pretrained encoder to 3D volume data.}
\label{main_fig}
\end{figure*}

\subsection{Frepa pretraining strategy}
\subsubsection{Frequency dual-component masking}
To enhance image representation of high-frequency information, we introduce the frequency dual-component masking strategy. This approach consists of two steps: high-frequency masking and low-frequency perturbation. By predicting image components of the masked frequencies, the image encoder can learn the high-frequency components and their distributions. Simultaneously, low-frequency perturbation prevents the encoder from overly relying on simple low-frequency components, instead encouraging it to utilize medium-frequency and high-frequency components for image reconstruction. This pretraining strategy can be seamlessly applied to both vanilla ViT, multi-scale ViTs, and convolutional networks without requiring changes to the framework (Fig.~\ref{main_fig}, a).
 
\textbf{High-frequency domain masking} Given a 2D image $x\in \mathbb{R} ^{H\times W}$, we can obtain its frequency spectrum $f\in \mathbb{R} ^{H\times W}$ via the 2D discrete centered Fourier transform:
\begin{equation}
f_{uv}=\sum_{h=0}^{H-1}{\sum_{w=0}^{W-1}{x_{wh}\cdot \left( -1 \right) ^{h+w}\cdot e^{-i2\pi \left( \frac{uh}{H}+\frac{vw}{W} \right)}}}.
\end{equation}
Here, the subscript $hw$ denotes the value of the image at the coordinate $(h, w)$, and similarly, the subscript $uv$ denotes the frequency spectrum at the coordinate $(u, v)$. Masking the low-frequency components can result in significant image distortion that is extremely challenging for reconstruction; therefore, we adopt a progressive masking strategy. Our goal is to make patch masking more likely to occur in the high-frequency components and less likely in the low-frequency components. To achieve this, we set the size of patches in the frequency spectrum to $16 \times 16$ and use an exponential decay function to control the masking probability (Fig.~\ref{main_fig},b). 
\begin{equation}
p_{ij}=1-\exp \left( -\frac{d_{ij}^{2}}{d_{c}^{2}} \right).
\label{distance}
\end{equation}
Here $p_{ij}$ means the masking probability of patch $ij$, and $d_{ij}$ is the Euclidean distance from the center of patch $ij$ to the center of the frequency spectrum. $d_{c}$ is the cut-off distance, and we set $d_{\boldsymbol{c}}=0.2\times \min \left( H, W \right)$ (Fig.~\ref{main_fig}, b). 

\textbf{Low-frequency perturbation} After the aforementioned stage, the high-frequency components are severely distorted. However, most low-frequency components remain unaffected, as $p_{ij}$ is much lower in the low-frequency domain. To prevent image restoration from relying solely on low-frequency components and instead promote understanding of high-frequency components, we introduce the low-frequency perturbation strategy. Specifically, we add intensity-attenuated, zero-mean noise to the masked frequency spectrum:
\begin{equation}
\begin{cases}
	\delta _{uv}\sim \mathcal{N} \left( 0, \sigma ^2 \right) \times \exp \left( -\frac{d_{uv}^{2}}{d_{c}^{2}} \right)\\
	\delta _{\frac{H}{2}, \frac{W}{2}}=0\\
\end{cases}
\label{perturbation}
\end{equation}
Here $\mathcal{N}$ is the Gaussian distribution, and $\sigma$ is the standard deviation. We set $\sigma=0.002\times \max \left( f \right)$ to make the root mean square error (RMSE) of perturbed images maintaining approximately the same with other image masking pretraining methods (\cite{mfm, simmim}). We also set the center of the noise to zero to ensure that the mean value of the image remains unchanged before and after the perturbation. Following these two stages, we perform the inverse Fourier transform on the corrupted frequency spectrum to obtain the degraded image.

\subsubsection{Equaled-histogram image-domain masking}
\begin{figure}[!t]
\centering
\centerline{\includegraphics[width=0.95\columnwidth]{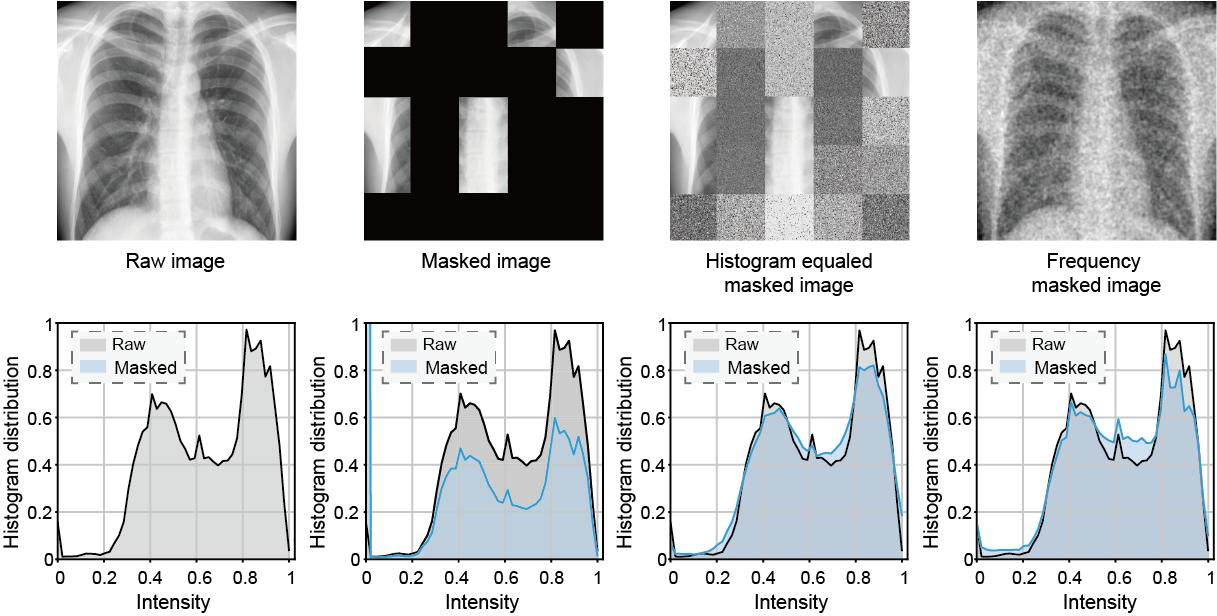}}
\caption{Histogram of different image masking strategies. Direct zero-masking will result in a severe shift of histogram distribution, while our proposed two masking strategies can preserve the raw distribution.}
\label{dist}
\end{figure}

Although advanced multi-scale ViTs and convolutional networks have demonstrated superior high-frequency capturing abilities compared to the vanilla ViT~\cite{hat, swin}, the success of MAE relies on the global property and asymmetric structure of the vanilla ViT, which enables the application of the self-attention mechanism to discrete, non-overlapping image patches~\cite{vit_success}. In contrast, multi-scale ViTs and convolutional networks typically incorporate local window operators, such as convolutional layers, which prevent them from this same capability. It has been shown that directly setting all masked patches to zero is inefficient~\cite{SparK}, as this approach significantly distorts the image histogram distribution (Fig.~\ref{dist}). Additionally, the convolutional operator can result in the ``mask pattern vanishing" issue~\cite{SparK}, causing inconsistent masking effects and ratios across different encoding layers.

To address this problem, we propose an innovative and straightforward image masking strategy, termed the histogram-equalized masking strategy. Specifically, we first uniformly sample 70$\%$ of the visible image patches, each with a size of $32 \times 32$. Unlike MAE~\cite{mae}, we do not discard the masked patches; instead, we replace them with random noise that has the same histogram distribution to the original patches. Denoting the one masked patch as $m_{ij}$, such an operation can be formulated as
\begin{equation}
    m_{ij}\gets \xi _{ij} \sim \mathcal{D} \left( m_{ij} \right),
\end{equation}
where $\xi _{ij}$ is the histogram-equalized noise, and $\mathcal{D} \left( \cdot \right)$ denotes the histogram distribution. Practically, such distributions can be approximated by a Gaussian distribution with the same mean and variance as the original patches. Images corrupted via this strategy can keep the histogram consistency before and after the masking, as shown in Fig.~\ref{dist}. Notably, this figure also shows that the frequency dual-component masking can also preserve the histogram distribution.

\subsection{Pretraining pipeline of Frepa}
Given a raw 2D image $x$, we randomly use the frequency-domain mask and the image-domain masking with equal probability to perform corruption and obtain the degraded image $\tilde{x}$. Our objective is to reconstruct $x$ from $\tilde{x}$ using an encoder-decoder network $N$:
\begin{equation}
N\left( \tilde{x} \right) \rightarrow x.
\end{equation}
We employ lightweight image decoders for the network, consisting of only a few convolutional layers. Compared to the Transformer decoder used in MAE~\cite{mae}, convolutional decoders offer significantly lower computational complexity~\cite{complexity}, particularly given that medical images often have relatively large dimensions (e.g., $512 \times 512$ in our experiments). Additionally, convolutional decoders demonstrate better capability in capturing local fine-grained details~\cite{hat}, aligning well with the objectives of our reconstruction task. To better guide the image encoder in leveraging high-frequency components and prevent key high-frequency details from being diluted by low-frequency components, we employ a high-frequency augmentation strategy. Specifically, we manually extract the high-frequency and edge features from the raw image via a filter operator $O$, and then concatenated it with the degraded image along the channel dimension. Then the pretraining process can be formulated as 
\begin{equation}
N\left[ \left( O\circ x \right) \oplus \tilde{x} \right] \rightarrow x.
\end{equation}
To prevent leakage of high-frequency components, the filter used cannot be a frequency-domain filter, such as exponential filtering; instead, filters or edge detection functions performed in the image domain are more appropriate. In our experiments, we use the Hessian filter~\cite{hessian} to accomplish this task. For simplicity, we use $N\left( \tilde{x} \right)$ instead of $N\left[ \left( O\circ x \right) \oplus \tilde{x} \right]$ to represent the reconstructed image in the following text. 

\subsubsection*{Loss function for reconstruction}
We utilize the RMSE to guide the reconstruction in the image domain.
\begin{equation}
\mathscr{L} _{\mathrm{RMSE}}=\sqrt{\frac{1}{H\times W}\left\| N\left( \tilde{x} \right) -x \right\| ^2}.
\end{equation}

However, this constraint may be insufficient to ensure similarity in image details and high-frequency components, as RMSE primarily focuses on overall pixel-wise intensity differences. Therefore, we additionally incorporate two loss functions, image gradient loss, and frequency-to-spatial loss, to better guide the reconstruction of fine-grained details and high-frequency components. The image gradient loss restricts the reconstruction of structures and details~\cite{gradient}.
\begin{equation}
\mathscr{L} _{\mathrm{Grad}}=\frac{1}{H\times W}\left\| \nabla N\left( \tilde{x} \right) -\nabla x \right\|,
\end{equation}
where $\nabla$ is the gradient operator. 

\begin{figure}[!t]
\centering
\centerline{\includegraphics[width=0.95\columnwidth]{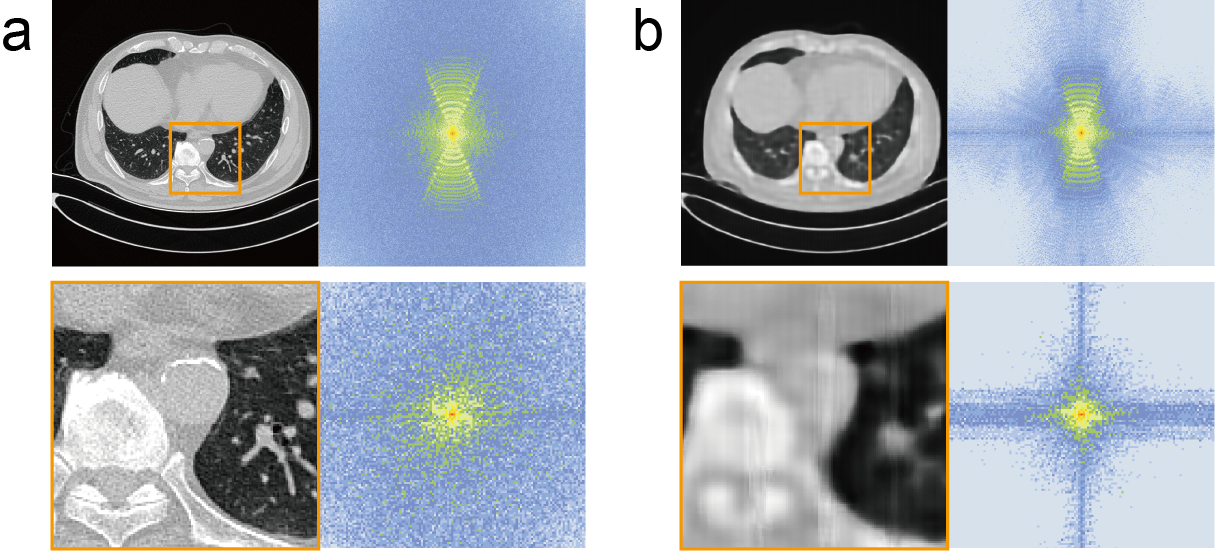}}
\caption{Comparison of the raw image (a) and the reconstructed image (b) optimized with FFL loss. The frequency spectrum is shown on the right side. The reconstructed images involve aliasing artifacts, which could be attributed to the over-fitting in the frequency domain. The image inside the orange box is zoomed in for better visualization.}
\label{artifact}
\end{figure}

To further enhance the representation of high-frequency components, we aim for the reconstructed images to achieve satisfactory alignment with the raw images in the frequency domain. Initially, we employ the commonly used focal frequency loss (FFL)~\cite{ffl}, specifically designed to enhance attention to high frequencies. However, in practice, we found that directly incorporating the FFL can result in aliasing artifacts in the reconstructed images (Fig.~\ref{artifact}), which cannot be ideal reconstruction outcomes. We tend to consider that these artifacts may arise from the degeneration of high-frequency information due to overfitting in the frequency domains~\cite{high_1}, as FFL could impose excessively strong constraints on high frequencies~\cite{usfm}. To address this issue, we propose a novel hierarchical frequency-to-spatial loss (HFL), which can be formulated as
\begin{equation}
\mathscr{L} _{\mathrm{HFL}}=\frac{1}{5}\sum_{k=1}^5{\frac{1}{H\times W}\left\| O_{h}^{k}\circ N\left( \tilde{x} \right) -O_{h}^{k}\circ x \right\|}.
\end{equation}
$O_{h}^{k}$ denotes the high-pass exponential filter, while a higher $k$ represents less preservation of the low-frequency components. In our implementation, the cut-off distance of the exponential filter is set to $\left\{ 0.1,0.2,0.3,0.4,0.5 \right\} \times \min \left( H, W \right) $, with $k$ ranging from 1 to 5. This hierarchical decomposition enables the optimization of high-frequency reconstruction in the image domain without directly imposing restrictions on the frequency spectrum, thereby preventing the formation of artifacts.

We also incorporate a training strategy to enhance the representation consistency, which encourages the image embeddings to remain consistent under different masking and perturbation samplings. Specifically, for a given input image $x$, we generate two degraded versions by applying distinct masking and perturbation configurations, $\left( m_1, \delta _1 \right)$ and $\left( m_2, \delta _2 \right)$, and the corresponding degraded images are denoted as $\tilde{x}^{m_1, \delta _1}$ and $\tilde{x}^{m_2, \delta _2}$, respectively. The consistency objective aims to make the embeddings of these two degraded images as similar as possible, and we use the Jensen-Shannon (JS) divergence to evaluate their distribution divergence:
\begin{equation}
\mathscr{L} _{\mathrm{Con}}=\mathrm{JS}\left[ E\left( \tilde{x}^{m_1, \delta _1} \right) , E\left( \tilde{x}^{m_2, \delta _2} \right) \right],
\end{equation}
where $E\left( \cdot \right) $ denotes the image encoder. The total loss functions can thus be formulated as follows:
\begin{equation}
\mathscr{L} _{\mathrm{total}}=\mathscr{L} _{\mathrm{RMSE}}+\lambda _1\mathscr{L} _{\mathrm{Grad}}+\lambda _2\mathscr{L} _{\mathrm{HFL}}+\lambda _3\mathscr{L} _{\mathrm{Con}}.
\end{equation}

In our experiments, we set $\left( \lambda _1, \lambda _2, \lambda _3 \right) =\left( 1, 1, 0.5 \right)$. For the images corrupted with the image-domain masking, $\mathscr{L} _{\mathrm{RMSE}}$ is only calculated on the masked patches, which is consistent with MAE~\cite{mae}; otherwise, the loss function is performed on the whole image. To enhance stability during the early stages of training, adversarial training is excluded for the first 5 epochs. Following this initial phase, the model is then trained using the previously described setting.

\subsection{Extending to 3D volume}
In clinical practice, many datasets are composed of 3D volumes and image sequences, such as computed tomography (CT) and magnetic resonance imaging (MRI) scans. The pretraining strategy can be readily adapted for 3D volume data or applied by transferring pretrained 2D encoders directly to 3D volumes (Fig.~\ref{main_fig}c).

\subsubsection{Adapt pretraining strategy for 3D volume}\label{pretrain_3D}
Given a 3D volume $v\in \mathbb{R} ^{H\times W \times N}$ consisting of $N$ stacked 2D images, similar masking strategies can be applied. For frequency dual-component masking, we first compute its 3D frequency spectrum $f\in \mathbb{R} ^{H\times W \times N}$ and apply high-frequency domain masking by modifying Eq.~\ref{distance} into a 3D manner
\begin{equation}
p_{ijk}=1-\exp \left( -\frac{d_{ijk}^{2}}{d_{c}^{2}} \right).
\label{distance_3D}
\end{equation}
Here $p_{ijk}$ is the masking probability of 3D patch $ijk$, while $d_{ijk}$ denotes the 3D Euclidean distance from the center of patch $ijk$ to the frequency spectrum center. Low-frequency perturbation can also be implemented by adjusting the Gaussian distribution in Eq.\ref{perturbation} to 3D distribution. In the equaled-histogram image-domain masking, we set the patch size to $16 \times 16 \times 16$~\cite{3dmae} and perform masking accordingly, 
\begin{equation}
    m_{ijk}\gets \xi _{ijk} \sim \mathcal{D} \left( m_{ijk} \right).
\end{equation}
Here $m_{ijk}$ is the masked patch in the 3D volume. The pretraining pipeline remains the same with the 2D operations. 

\subsubsection{Transfer pretrained 2D encoder to 3D volumes}
Although Frepa can be directly adapted for 3D volume data, re-pretraining can cost extensive computational resources. Here we propose a method that can transfer pretrained 2D encoders directly to 3D volumes without re-training. Specifically, given a volume data $v\in \mathbb{R} ^{H\times W \times N}$, we first apply the pretrained 2D encoder to each stacked image to obtain embeddings $e\in \mathbb{R} ^{\left( N\times n_C \right) \times n_H\times n_H}$ with feature channels $n_C$ and embedding size $ n_H \times n_W$. To preserve the spatial information within each image, fixed sinusoidal positional embedding $ p_{\text{intra}} $ is added to the rearranged embeddings as $ e + p_{\text{intra}}$. The resulting embeddings are then processed by an intra-image spatial transformer block, which employs a multi-head self-attention mechanism and a feedforward network (following the standard transformer architecture~\cite{transformer_block}) to capture spatial relationships within each individual image. This step results in output embeddings $ e_{\text{intra}} \in \mathbb{R}^{(N \times n_C) \times (n_H \times n_W) \times n_d}$, where $ n_d $ denotes the token dimension.

The embeddings $ e_{\text{intra}}$ are subsequently reshaped to $ \mathbb{R}^{(n_H \times n_W) \times (N \times n_C) \times n_d} $, enabling the transformer to focus on relationships between corresponding patches across the $N$ stacked images. To encode the relative positions of slices within the stack, an inter-image positional embedding $ p_{\text{inter}} $ is added to the rearranged embeddings. The resulting representation is then processed by an inter-image transformer block (same architecture as the intra-image transformer block), which captures contextual relationships between the stacked images by attending to corresponding patches across slices. The intra- and inter-image transformer blocks are applied iteratively (three iterations in the current implementation) to progressively integrate intra-image and inter-image information, thereby refining the representation of both local and global patterns. Finally, the output embeddings are passed to a decoder head tailored for various downstream tasks. During the training phase, the parameters of the pretrained 2D encoder are frozen, while all parameters within the transformer blocks are optimized.

\section{Experiments and results}
\begin{table}[!h]
\resizebox{0.95\linewidth}{!}{
\begin{tabular}{|c|c|c|}
\hline
\rowcolor[HTML]{ECF4FF} 
\textbf{Modality}                         & \textbf{Datasets}               & \textbf{Number of   images} \\ \hline
                                          & DeepLesion~\cite{deeplesion}                      & 928,020                     \\ \cline{2-3} 
                                          & Colonography dataset~\cite{Colonography}             & 660,207                     \\ \cline{2-3} 
                                          & MSD dataset~\cite{msd}     
                                                  & 147,796                     \\ \cline{2-3} 
                                          & CQ500~\cite{CQ500}                           & 72,282                      \\ \cline{2-3} 
\multirow{-5}{*}{CT (1,860,192)}        & TCIA HCC-TACE~\cite{HCC}                    & 51,887                      \\ \hline
                                          & UCSF-PDGM~\cite{UCSF-PDGM}                        & 11,523                      \\ \cline{2-3} 
                                          & BraTS2021~\cite{BraTS2021}                       & 79,112                      \\ \cline{2-3} 
                                          & Duke-Breast-Cancer-MRI~\cite{Duke-breast}          & 773,253                     \\ \cline{2-3} 
                                          & ACRIN-6698~\cite{ACRIN}                           & 891,786                     \\ \cline{2-3} 
\multirow{-5}{*}{MRI (1,773,510)}       & Colorectal-Liver-Metastases~\cite{Colorectal-Liver-Metastases}     & 17,836                      \\ \hline
                                          & CheXpert~\cite{chexpert}                       & 224,316                     \\ \cline{2-3} 
                                          & MIMIC-CXR~\cite{mimic-cxr}                       & 371,920                     \\ \cline{2-3} 
                                          & NIH Chest~\cite{NIH_Chest}                       & 112,120                     \\ \cline{2-3} 
                                          & COVID-19 Radiography~\cite{COVID-19-Radiography}   & 21,165                      \\ \cline{2-3} 
\multirow{-5}{*}{X-ray (730,373)}       & COVIDGR~\cite{COVIDGR}                         & 852                         \\ \hline
                                          & Breast-Lesions-USG~\cite{Breast-Lesions}                          & 522                         \\ \cline{2-3} 
                                          & Fetal Plane~\cite{fetal}                    & 12,400                      \\ \cline{2-3} 
                                          & Muscle fascicle length tracking~\cite{Muscle} & 42,000                      \\ \cline{2-3} 
                                          & TN3K~\cite{TN3K}                            & 3,493                       \\ \cline{2-3} 
                                          & EBUS~\cite{EBUS}                            & 1,000                       \\ \cline{2-3} 
\multirow{-6}{*}{Ultrasound   (62,325)} & Micro-Ultrasound Prostate~\cite{Micro-Ultrasound}       & 2,910                       \\ \hline
                                          & OCT 2017~\cite{oct2017}              & 84,484                      \\ \cline{2-3} 
                                          & OCT classification~\cite{OCT_classification}              & 66,075                      \\ \cline{2-3} 
                                          & OCTDL~\cite{OPCTDL}                           & 2064                        \\ \cline{2-3} 
                                          & OLIVES~\cite{OLIVES}                          & 33,533                      \\ \cline{2-3} 
\multirow{-4}{*}{OCT (210,167)}         & RetinalOCT~\cite{RentialOCT}                      & 24,000                      \\ \hline
                                          & ODIR-5K~\cite{ODIR-5K}                         & 5,000                       \\ \cline{2-3} 
                                          & Aptos2019~\cite{Aptos2019}                       & 3,662                       \\ \cline{2-3} 
\multirow{-3}{*}{Retina (72,081)}       & Diabetic Retinopathy Detection~\cite{Diabetic_Retinopathy}  & 63,419                      \\ \hline
                                          & ISIC2019~\cite{isic}                            & 77,245                      \\ \cline{2-3} 
\multirow{-2}{*}{Dermoscopy (98,397)}  & Private dataset $\#$                 & 21,152                      \\ \hline
\end{tabular}}
\caption{Summary of the dataset for model pretraining. The total image numbers of each modality are marked in brackets. "$\#$" denotes the in-house datasets.}
\label{dataset}
\end{table}

\begin{table}[]
\resizebox{0.95\linewidth}{!}{
\begin{tabular}{|c|c|c|}
\hline
\rowcolor[HTML]{FFFFC7} 
\textbf{Modality}                    & \textbf{Datasets}     & \textbf{Downstream} \\ \hline
                                     & Tianchi cancer~\cite{tianchi_cancer} $\dagger$          & Classification      \\ \cline{2-3} 
                                     & Lung tumor~\cite{nodule_1}  $\dagger$          & Detection      \\ \cline{2-3}
                                     & COPD$\#$  $\dagger$               & Classification      \\ \cline{2-3} 
                                     & Pneumonia$\#$  $\dagger$            & Classification      \\ \cline{2-3} 
                                     & Luna16 Lung Nodule~\cite{luna16}  $\dagger$   & Detection           \\ \cline{2-3} 
                                     & Lung$\#$ $\dagger$               & Segmentation        \\ \cline{2-3} 
                                     & Heart$\#$ $\dagger$             & Segmentation        \\ \cline{2-3} 
                                     & Pneumonia infection~\cite{penimonia}  $\dagger$ & Segmentation        \\ \cline{2-3} 
                                     & Pulmonary artery~\cite{av} $\dagger$      & Segmentation        \\ \cline{2-3} 
                                     & Pulmonary vein~\cite{av} $\dagger$        & Segmentation        \\ \cline{2-3} 
                                     & Abdomen1K-Liver~\cite{abdo_1K} $\dagger$       & Segmentation        \\ \cline{2-3} 
                                     & Abdomen1K-Kidney~\cite{abdo_1K} $\dagger$      & Segmentation        \\ \cline{2-3} 
\multirow{-13}{*}{CT}                & Abdomen1K-Spleen~\cite{abdo_1K} $\dagger$      & Segmentation        \\ \hline
                                     & AMOS22~\cite{amos} $\dagger$                & Segmentation        \\ \cline{2-3} 
\multirow{-2}{*}{MRI}                & ACDC~\cite{acdc} $\dagger$                  & Segmentation        \\ \hline
                                     & COVID19 Radiography~\cite{COVID-19-Radiography}   & Classification      \\ \cline{2-3} 
                                     & Pneumonia MNIST~\cite{medmnist}  $\dagger$      & Classification      \\ \cline{2-3} 
\multirow{-3}{*}{Xray}               & COVID19 PA Dataset~\cite{covid_2}  $\dagger$   & Classification      \\ \hline                
                                     & Fetal Plane~\cite{fetal}           & Classification      \\ \cline{2-3} 
\multirow{-2}{*}{Ultrasound}         & Breast MNIST~\cite{medmnist} $\dagger$         & Classification      \\ \hline
                                     & OCT 17~\cite{oct2017}              & Classification      \\ \cline{2-3} 
\multirow{-2}{*}{OCT}                & OCT MNIST~\cite{medmnist}  $\dagger$            & Classification      \\ \hline
                                     & Aptos2019~\cite{Aptos2019}             & Classification      \\ \cline{2-3} 
                                     & ISIC2019~\cite{isic}              & Classification      \\ \cline{2-3} 
                                     & HRF~\cite{hrf} $\dagger$                   & Segmentation        \\ \cline{2-3} 
                                     & DRIVE~\cite{drive} $\dagger$                & Segmentation        \\ \cline{2-3} 
\multirow{-5}{*}{Retina}             & CHASEDB1~\cite{CHAOS} $\dagger$             & Segmentation        \\ \hline
                                     & OCTA artery~\cite{octa} $\dagger$           & Segmentation        \\ \cline{2-3} 
                                     & OCTA vein~\cite{octa} $\dagger$             & Segmentation        \\ \cline{2-3} 
                                     & OCT artery~\cite{octa} $\dagger$            & Segmentation        \\ \cline{2-3} 
\multirow{-4}{*}{OCT reconstruction} & OCT vein~\cite{octa} $\dagger$              & Segmentation        \\ \hline
EM                                   & CREMI~\cite{CREMI} $\dagger$                 & Segmentation        \\ \hline
\end{tabular}}
\caption{Summary of the dataset for downstream task evaluation. "$\#$" denotes the in-house datasets, and "$\dagger$" means the external dataset, which is not seen in the pretraining phase.}
\label{downstream}
\end{table}

\begin{table*}[!h]
\centering
\renewcommand{\arraystretch}{1.1}
\resizebox{1.0\textwidth}{!}{
\begin{tabular}{|c|c|c|c|c|c|c|c|c|c|c|}
\hline
\rowcolor[HTML]{EFEFEF} 
Modality                       & Dataset                               & Metric    & MAE         & MFM         & CLIP                 & MedSAM      & MedDr       & Frepa+ViT            & Frepa+SwinT          & Frepa+ConvNeXt       \\ \hline
                               &                                       & ACC       & 0.941$\pm$0.023 & 0.903$\pm$0.016 & \textbf{0.962$\pm$0.013} & 0.951$\pm$0.026 & 0.887$\pm$0.049 & \underline{0.959$\pm$0.020}    & 0.959$\pm$0.022          & 0.959$\pm$0.015          \\ \cline{3-11} 
                               & \multirow{-2}{*}{Tianchi   Cancer}    & AUC       & 0.982$\pm$0.013 & 0.973$\pm$0.010  & \textbf{0.992$\pm$0.004} & 0.986$\pm$0.007 & 0.970$\pm$0.013 & 0.985$\pm$0.012          & \underline{0.987$\pm$0.012}    & 0.985$\pm$0.008          \\ \cline{2-11} 
                               &                                       & Detection & 0.916$\pm$0.012 & 0.905$\pm$0.013 & 0.638$\pm$0.072          & 0.894$\pm$0.014 & 0.712$\pm$0.035 & \textbf{0.954$\pm$0.009} & \underline{0.943$\pm$0.006}    & 0.937$\pm$0.006          \\ \cline{3-11} 
\multirow{-4}{*}{CT}           & \multirow{-2}{*}{Lung tumor}         & IoU       & 0.601$\pm$0.067 & 0.613$\pm$0.072 & 0.275$\pm$0.148          & 0.572$\pm$0.099 & 0.455$\pm$0.142 & 0.662$\pm$0.061          & \textbf{0.684$\pm$0.086} & \underline{0.677$\pm$0.070}     \\ \hline
                               &                                       & ACC       & 0.965$\pm$0.010  & 0.962$\pm$0.008 & \textbf{0.983$\pm$0.004} & 0.970$\pm$0.010  & 0.968$\pm$0.004 & \underline{0.976$\pm$0.010}     & 0.972$\pm$0.007          & 0.974$\pm$0.008          \\ \cline{3-11} 
                               & \multirow{-2}{*}{COVID19 Radiography} & AUC       & 0.997$\pm$0.001 & 0.996$\pm$0.002 & \textbf{0.998$\pm$0.001} & 0.997$\pm$0.001 & 0.996$\pm$0.001 & 0.998$\pm$0.001          & \underline{0.998$\pm$0.001}    & 0.998$\pm$0.001          \\ \cline{2-11} 
                               &                                       & ACC       & 0.975$\pm$0.003 & 0.969$\pm$0.003 & \underline{0.975$\pm$0.002}    & 0.969$\pm$0.006 & 0.962$\pm$0.004 & \textbf{0.977$\pm$0.004} & 0.974$\pm$0.002          & 0.974$\pm$0.004          \\ \cline{3-11} 
                               & \multirow{-2}{*}{Pneumonia MNIST}     & AUC       & 0.995$\pm$0.002 & 0.994$\pm$0.002 & \textbf{0.996$\pm$0.001} & 0.993$\pm$0.003 & 0.991$\pm$0.003 & \underline{0.995$\pm$0.002}    & 0.994$\pm$0.002          & 0.994$\pm$0.002          \\ \cline{2-11} 
                               &                                       & ACC       & 0.940$\pm$0.006 & 0.935$\pm$0.014 & \textbf{0.951$\pm$0.006} & 0.944$\pm$0.011 & 0.940$\pm$0.015 & \underline{0.944$\pm$0.009}    & 0.941$\pm$0.007          & 0.939$\pm$0.012          \\ \cline{3-11} 
\multirow{-6}{*}{X-Ray}        & \multirow{-2}{*}{COVID19 PA Dataset}  & AUC       & 0.981$\pm$0.002 & 0.981$\pm$0.007 & \underline{0.987$\pm$0.004}    & 0.982$\pm$0.003 & 0.983$\pm$0.007 & \underline{0.984$\pm$0.003}    & 0.982$\pm$0.007          & 0.980$\pm$0.007          \\ \hline
                               &                                       & ACC       & 0.859$\pm$0.008 & 0.883$\pm$0.009 & \textbf{0.903$\pm$0.006} & 0.890$\pm$0.009  & 0.819$\pm$0.012 & 0.901$\pm$0.004          & 0.896$\pm$0.005          & \underline{0.901$\pm$0.004}    \\ \cline{3-11} 
                               & \multirow{-2}{*}{OCT17}               & AUC       & 0.963$\pm$0.021 & 0.970$\pm$0.010 & \underline{0.985$\pm$0.013}    & 0.968$\pm$0.024 & 0.942$\pm$0.019 & \textbf{0.989$\pm$0.019} & 0.985$\pm$0.022          & 0.982$\pm$0.015          \\ \cline{2-11} 
                               &                                       & ACC       & 0.899$\pm$0.005 & 0.891$\pm$0.004 & 0.915$\pm$0.005          & 0.882$\pm$0.005 & 0.844$\pm$0.006 & \textbf{0.927$\pm$0.004} & \underline{0.917$\pm$0.006}    & 0.912$\pm$0.005          \\ \cline{3-11} 
\multirow{-4}{*}{OCT}          & \multirow{-2}{*}{OCT MNIST}           & AUC       & 0.963$\pm$0.004 & 0.957$\pm$0.002 & \underline{0.972$\pm$0.002}    & 0.954$\pm$0.003 & 0.939$\pm$0.003 & \textbf{0.975$\pm$0.002} & 0.971$\pm$0.002          & 0.969$\pm$0.002          \\ \hline
                               &                                       & ACC       & 0.934$\pm$0.001 & 0.930$\pm$0.003 & 0.928$\pm$0.003          & 0.924$\pm$0.003 & 0.860$\pm$0.004 & \textbf{0.937$\pm$0.002} & 0.931$\pm$0.004          & \underline{0.935$\pm$0.001}    \\ \cline{3-11} 
                               & \multirow{-2}{*}{Fetal Ultra}         & AUC       & 0.965$\pm$0.004 & 0.968$\pm$0.004 & 0.970$\pm$0.003          & 0.964$\pm$0.002 & 0.910$\pm$0.008 & \underline{0.972$\pm$0.004}    & \textbf{0.973$\pm$0.003} & 0.972$\pm$0.005          \\ \cline{2-11} 
                               &                                       & ACC       & 0.869$\pm$0.017 & 0.843$\pm$0.035 & \textbf{0.887$\pm$0.017} & 0.857$\pm$0.029 & 0.830$\pm$0.050 & \underline{0.873$\pm$0.032}    & 0.866$\pm$0.044          & 0.861$\pm$0.011          \\ \cline{3-11} 
\multirow{-4}{*}{Ultrasound}   & \multirow{-2}{*}{Breast MNIST}        & AUC       & 0.839$\pm$0.041 & 0.820$\pm$0.054 & \textbf{0.900$\pm$0.036} & 0.857$\pm$0.011 & 0.822$\pm$0.054 & \underline{0.870$\pm$0.027}    & 0.859$\pm$0.024          & 0.859$\pm$0.034          \\ \hline
                               &                                       & ACC       & 0.717$\pm$0.023 & 0.736$\pm$0.019 & 0.707$\pm$0.020          & 0.740$\pm$0.013 & 0.689$\pm$0.019 & \underline{0.761$\pm$0.011}    & \textbf{0.764$\pm$0.012} & 0.754$\pm$0.011          \\ \cline{3-11} 
\multirow{-2}{*}{Retina}       & \multirow{-2}{*}{Aptos2019}           & AUC       & 0.873$\pm$0.013 & 0.890$\pm$0.009 & 0.882$\pm$0.019          & 0.886$\pm$0.020 & 0.845$\pm$0.023 & \textbf{0.904$\pm$0.007} & 0.897$\pm$0.009          & \underline{0.899$\pm$0.008}    \\ \hline
                               &                                       & ACC       & 0.719$\pm$0.005 & 0.714$\pm$0.011 & 0.712$\pm$0.006          & 0.716$\pm$0.007 & 0.637$\pm$0.002 & 0.738$\pm$0.004          & \underline{0.742$\pm$0.001}    & \textbf{0.744$\pm$0.003} \\ \cline{3-11} 
\multirow{-2}{*}{Dermatoscope} & \multirow{-2}{*}{ISIC2019}            & AUC       & 0.918$\pm$0.003 & 0.915$\pm$0.004 & 0.907$\pm$0.005          & 0.914$\pm$0.002 & 0.856$\pm$0.004 & 0.932$\pm$0.005          & \textbf{0.935$\pm$0.003} & \underline{0.932$\pm$0.003}    \\ \hline
\end{tabular}}
\caption{Performance on 2D image classification and detection tasks. "$\dagger$" means the external dataset, which is not seen in the pretraining phase. "ACC"=Accuracy, "AUC"=Area under the curve. “IoU”=Intersection over union. \textbf{Bold} indicates the best results, and \underline{underline} indicates the second best results.}
\label{class_2D}
\end{table*}

\begin{table*}[!h]
\centering
\resizebox{\textwidth}{!}{
\begin{tabular}{|c|c|c|c|c|c|c|c|c|c|c|}
\hline
\rowcolor[HTML]{EFEFEF} 
Modality             & Dataset          & Metric & MAE                          & MFM                          & CLIP                  & MedSAM               & MedDr               & Frepa+ViT           & Frepa+SwinT         & Frepa+ConvNeXt      \\ \hline
                     &                  & ACC    & 0.853$\pm$0.037              & 0.816$\pm$0.106              & 0.869$\pm$0.024       & 0.800$\pm$0.053      & 0.656$\pm$0.035     & \textbf{0.896$\pm$0.017} & 0.851$\pm$0.076 & \underline{0.875$\pm$0.028} \\ \cline{3-11} 
                     & \multirow{-2}{*}{COPD}             & AUC    & 0.942$\pm$0.062              & 0.926$\pm$0.074              & 0.936$\pm$0.061       & 0.920$\pm$0.044      & 0.854$\pm$0.075     & \textbf{0.961$\pm$0.031} & 0.956$\pm$0.035 & \underline{0.960$\pm$0.046} \\ \cline{2-11} 
                     &                  & ACC    & 0.860$\pm$0.028              & 0.868$\pm$0.054              & 0.876$\pm$0.026       & \underline{0.888$\pm$0.052} & 0.804$\pm$0.041     & 0.884$\pm$0.030          & 0.864$\pm$0.017 & \textbf{0.888$\pm$0.048} \\ \cline{3-11} 
                     & \multirow{-2}{*}{Pneumonia}        & AUC    & 0.928$\pm$0.061              & 0.917$\pm$0.052              & \underline{0.935$\pm$0.050} & 0.930$\pm$0.074      & 0.865$\pm$0.095     & \textbf{0.938$\pm$0.065} & 0.929$\pm$0.059 & 0.935$\pm$0.063 \\ \cline{2-11} 
                     &                  & Sen    & 0.846$\pm$0.066              & 0.834$\pm$0.076              & 0.697$\pm$0.146       & 0.875$\pm$0.076      & 0.816$\pm$0.071     & 0.873$\pm$0.087          & \underline{0.885$\pm$0.091} & \textbf{0.889$\pm$0.078} \\ \cline{3-11} 
\multirow{-6}{*}{CT} & \multirow{-2}{*}{Luna16}           & Pre    & 0.873$\pm$0.058              & 0.823$\pm$0.059              & 0.691$\pm$0.125       & \textbf{0.893$\pm$0.068} & 0.829$\pm$0.059     & 0.879$\pm$0.057          & \underline{0.885$\pm$0.099} & 0.877$\pm$0.088 \\ \hline
\end{tabular}
}
\caption{Performance on 3D image classification and detection tasks. "$\dagger$" means the external dataset, which is not seen in the pretraining phase. “COPD”=Chronic obstructive pulmonary disease. "ACC"=Accuracy, "AUC"=Area under the curve. “Sen”=Sensitivity, "Pre"=Precision. \textbf{Bold} indicates the best results, and \underline{underline} indicates the second best results.}

\label{class_3D}
\end{table*}

\begin{table*}[!h]
\centering
\resizebox{0.8\textwidth}{!}{
\begin{tabular}{|c|c|c|c|c|c|c|}
\hline
\rowcolor[HTML]{EFEFEF} 
Modality             & Dataset                       & Metric    & 3D MAE               & 3DFrepa+ViT          & 3DFrepa+SwinT        & 3DFrepa+ConvNeXt     \\ \hline
                     &                               & ACC       & 0.872$\pm$0.028          & \underline{0.893$\pm$0.021}    & 0.888$\pm$0.024          & \textbf{0.896$\pm$0.032} \\ \cline{3-7} 
                     & \multirow{-2}{*}{COPD}      & AUC       & 0.954$\pm$0.05           & \underline{0.967$\pm$0.064}    & 0.962$\pm$0.043          & \textbf{0.967$\pm$0.056} \\ \cline{2-7} 
                     &                               & ACC       & 0.880$\pm$0.042          & 0.884$\pm$0.036          & \textbf{0.900$\pm$0.032} & \underline{0.900$\pm$0.037}    \\ \cline{3-7} 
                     & \multirow{-2}{*}{Pneumonia} & AUC       & \textbf{0.933$\pm$0.058} & 0.932$\pm$0.057          & 0.933$\pm$0.066          & \textbf{0.941$\pm$0.062} \\ \cline{2-7} 
                     &                               & Sen       & 0.893$\pm$0.048          & \underline{0.901$\pm$0.042}    & \textbf{0.905$\pm$0.067} & 0.891$\pm$0.057          \\ \cline{3-7} 
\multirow{-6}{*}{CT} & \multirow{-2}{*}{Luna16}      & Pre       & 0.888$\pm$0.035          & \textbf{0.924$\pm$0.031} & 0.908$\pm$0.048           & \underline{0.911$\pm$0.046}    \\ \hline
\end{tabular}
}
\caption{Performance on 3D image classification and detection tasks with 3D pretrained methods. "$\dagger$" means the external dataset, which is not seen in the pretraining phase. “COPD”=Chronic obstructive pulmonary disease. "ACC"=Accuracy, "AUC"=Area under the curve. “Sen”=Sensitivity, "Pre"=Precision. \textbf{Bold} indicates the best results, and \underline{underline} indicates the second best results.}

\label{3Dclass_3D}
\end{table*}

\begin{figure*}[!t]
\centerline{\includegraphics[width=\textwidth]{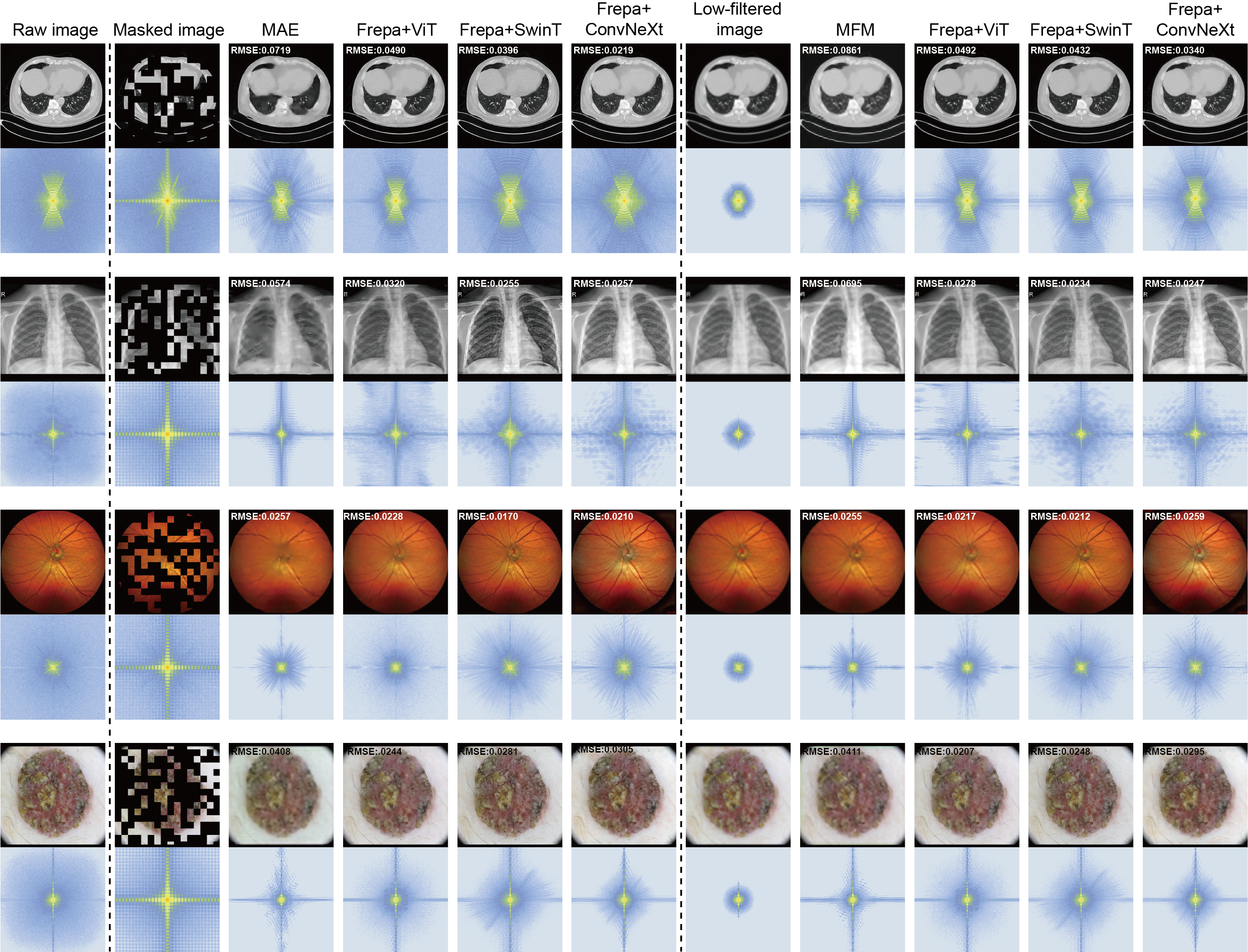}}
\caption{Example results of reconstructed images on external datasets. The images are corrupted by random masking and low-frequency filtering, respectively. Notably, such low-frequency filtered images are not seen during the training phases of Frepa. We visualize both the images and their frequency spectrum. RMAE is shown in the upper left corner of each image. Zoom in to an appropriate size for better viewing of the images.}

\label{recons}
\end{figure*}

\subsection{Implementation details}\label{ex_1}
\subsubsection{Pretrained on 2D images}
We conducted experiments on three popular architectures on 2D images: vanilla ViT-B~\cite{vit} (86.9M parameters), SwinT-B~\cite{swin} (86.9M parameters), and ConvNeXt~\cite{convnext} (feature channel=128, 87.5M parameters) as the image encoders. The image decoder for reconstruction consisted of four convolutional layers, while the classification head was composed of a multilayer perception. The models were pretrained on a comprehensive dataset, including modalities of CT, MRI, X-ray, ultrasound, optical coherence tomography (OCT, projection image), retina imaging, and dermatoscopy. Table~\ref {dataset} summarizes the datasets employed for model pretraining. For CT images, we initially normalized the Hounsfield units using typical window width and level values ([-1000, 400] for the lung window, [-160, 240] for the abdomen window, [-30, 90] for the brain window, and [-300, 1200] for the bone window) and then rescaled them to [0, 1]. For other imaging modalities, we clipped the intensity values to the range between the 0.5th and 99.5th percentiles before rescaling them to [0, 1]. All images were resized to $512\times 512 \times 3$ using bi-cubic interpolation and zero-padding. For single-channel grayscale images, the image was duplicated along the channel dimension to maintain consistency. The model was trained using the Adam optimizer with a learning rate of $1\times10^{-4}$. We employed image augmentation techniques including random flipping, random rotation, random affine transformations, and random histogram shifting during the pretraining phase. The models were trained on a Linux workstation with two A100 (80G) GPUs for 100 epochs.  

We compared Frepa with other pretrained and task-specific foundation models, including MAE~\cite{mae}, MFM~\cite{mfm} (a self-supervised strategy performed in the frequency domain), MedSAM~\cite{sam_1}, and CLIP~\cite{clip}. The backbones for MAE and MFM were ViT-B, consistent with Frepa, and were pretrained on our collected dataset. The training of CLIP followed the method proposed in \cite{meddr}. For MedSAM, we used the model and checkpoint published in \cite{sam_1} without modification. We also included one large visual-language model, MedDr~\cite{meddr} (40 billion parameters) for the comparison of classification tasks. All encoders were used solely for image encoding, with identical decoder structures applied across models. During the downstream task training phase, the parameters of all model encoders were kept fixed, and only the decoders were trained and optimized.

\subsubsection{Pretrained on 3D volumes}
We also conducted model pretraining experiments directly on 3D volumes, with the method described in Section~\ref{pretrain_3D}. Specifically, we included the same three backbones in their 3D versions: ViT (dim=1024, 32.0M parameters), SwinT (feature size=36, 35.1M parameters), and 3D ConvNeXt (MedNeXt~\cite{mednext}, feature channel=24, 34.8M parameters). The models were pretrained on all CT and MRI volumes from our established dataset (Table~\ref{dataset}). Following standard practices for handling 3D volumes, we randomly cropped the data into $128\times128\times128$ sub-volume cubes to address GPU memory limitations. This cube size was consistently applied during training, validation, and testing. The input channel was set to 1, as both CT and MRI modalities consist of grayscale images. We compared the 3D-pretrained Frepa model with 3D MAE~\cite{3dmae}, which directly extends the MAE strategy to 3D data. All other experimental settings were kept consistent with those used in the 2D experiments for a fair comparison.

\subsection{Experiment settings for downstream tasks}
We froze the encoders and fine-tuned only the task heads for downstream tasks. To ensure a fair comparison, we employed consistent training strategies and decoder architectures across all foundation models. The loss functions used were Cross Entropy loss for classification, Dice loss for segmentation, and GIoU loss~\cite{gIoU} for detection tasks. The datasets were randomly divided into training, validation, and testing sets in a 7:1:2 ratio. For the classification and detection tasks, we employed the five-fold training strategies to ensure robust evaluation and improve evaluation generalization. Table \ref{downstream} summarizes the datasets used for downstream task evaluation, respectively. In the tables, "$\#$" indicates in-house datasets, while "$\dagger$" denotes external datasets not utilized during the pretraining phase.

For classification evaluation, we utilized classification accuracy (ACC) and the area under the curve (AUC) metrics. Segmentation performance was assessed using the DSC. Detection ratio and IoU were used to evaluate cancer detection tasks, while sensitivity and precision were employed for the evaluation of the Luna16 lung nodule detection. This approach aligns with the metrics used in \cite{luna16} and \cite{luna16_sota}, with all calculations strictly following the methods outlined in their papers. Notably, we incorporated two modalities—OCT reconstruction data and electron microscopy (EM) data—for segmentation tasks. These modalities were not included in the pretraining phases, thereby evaluating the generalizability to real-world data. 

During the training phase for retinal vessel segmentation, individual datasets provide only a limited amount of training data (e.g., 45 images in HRF~\cite{hrf}, 20 images in DRIVE~\cite{drive}, and 28 images in the CHASEDB~\cite{CHAOS} dataset). To address this limitation, we combined these three datasets together for training while reporting the prediction results separately for each dataset.

\subsection{Visualization for reconstruction}
We present several visualizations of reconstruction results on external datasets. Specifically, we compare reconstruction for image masking against MAE and reconstruction for frequency masking, with low-pass filtering, against MFM. As shown in Fig.~\ref{recons}, Frepa demonstrates the highest fidelity in reconstructing original images, with exceptional detail information. In contrast, images generated by other autoencoders appear relatively blurred, lacking sufficient detail in textures and edges. These results are further corroborated by the frequency spectrum and RMSE metrics of each image. The superior reconstruction quality and clear detail retention indicate that Frepa effectively captures medium- and high-frequency information. 

\subsection{Comparison with previous methods}
\subsubsection{Classification and detection} 
Tables~\ref{class_2D}, ~\ref{class_3D}, and ~\ref{3Dclass_3D} present the results of classification and detection on 2D images and 3D volumes, respectively. It shows that Frepa, when implemented with both ViT, SwinT, or ConvNeXt, overall outperforms MAE and MAF. Specifically, Frepa exhibits an average improvement on ACC of 2.3$\%$ and 2.4$\%$ over MAE, and 3.0$\%$ and 3.1$\%$ over MFM, respectively. AUC metric also shows a similar conclusion. Furthermore, when compared to the classification-specific foundation model, CLIP, Frepa demonstrates comparable performance across most tasks and excels in certain areas such as the classification of chest cancer sub-types and pneumonia sub-types. Additionally, it is noteworthy that in fine-grained tasks, such as ISIC2019 fine-grained classification and lung tumor detection, Frepa achieves more substantial improvements, attributed to its superior capability of capturing high-frequency details. Specifically, Frepa achieves a 2.5$\%$ improvement in accuracy and a 1.7$\%$ increase in AUC on the ISIC2019 dataset, while it demonstrates a 4.0$\%$ increase in lung tumor detection ratio and a 7$\%$ improvement in IoU, as shown in Table~\ref{class_2D}, compared to the state-of-the-art (SOTA) results. Collectively, in classification tasks reliant on global-level information, Frepa achieves performance comparable to CLIP and MAE, affirming its robust global-level representation. Moreover, for tasks requiring fine-grained information, Frepa consistently outperforms, underscoring its effectiveness in capturing detailed, high-frequency information.


\subsubsection{Segmentation} 
\begin{figure*}[!t]
\centerline{\includegraphics[width=0.8\textwidth]{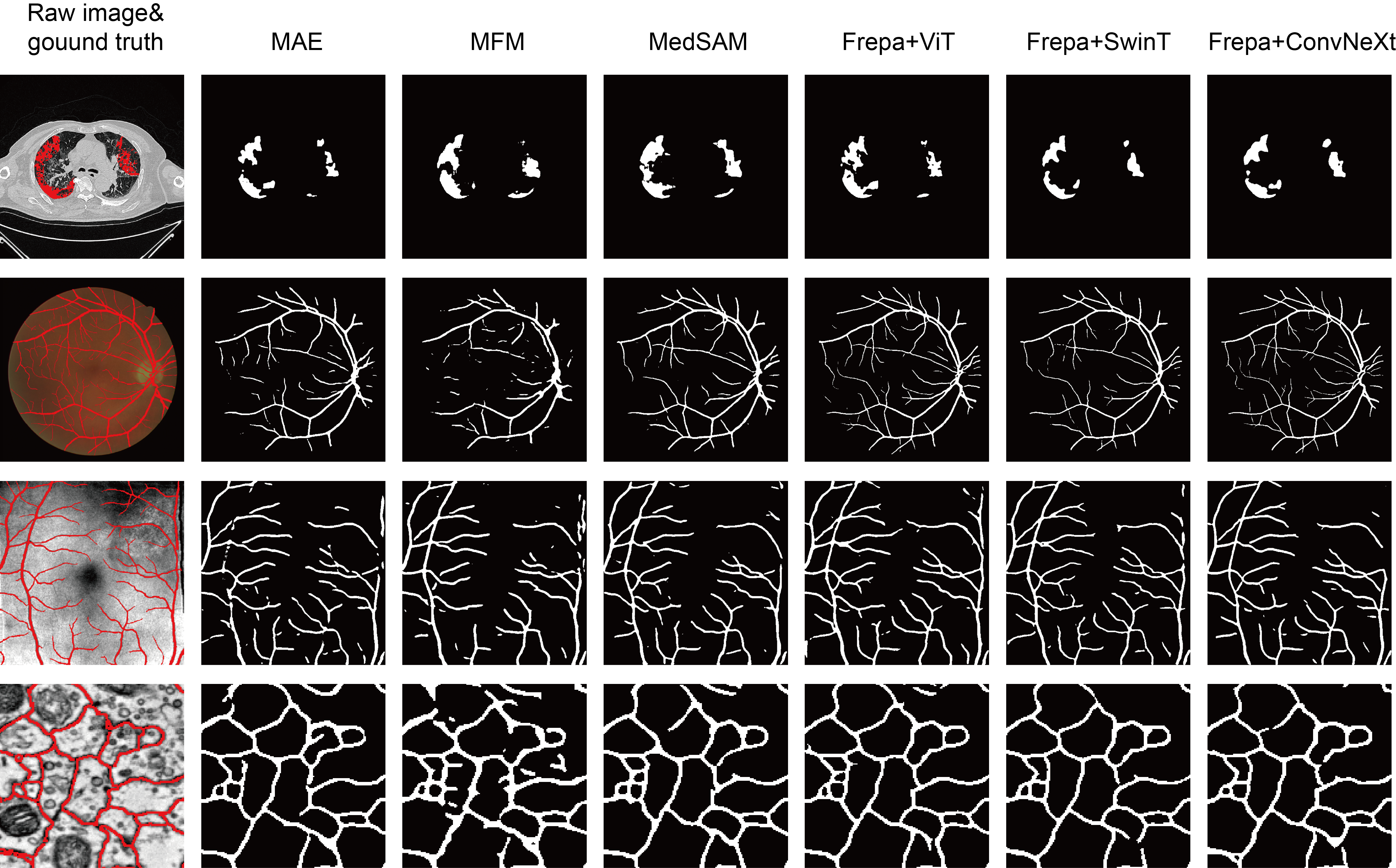}}
\caption{Visualization of segmentation results of pneumonia, retinal vessels, OCTA reconstruction vessels, and intercellular gaps, which all represent fine-grained detailed features. The ground truth segmentation is marked with red color. The segmentation results achieved by Frepa-pretrained models present higher topological structure accuracy and better fine-grained detail preservation.}
\label{seg}
\end{figure*}

Tables \ref{seg_2D}, \ref{seg_3D}, and \ref{3Dseg_3D} present the segmentation results on 2D images and 3D volumes, respectively. These results encompass both large-scale anatomical segmentations, (lung, heart, and pneumonia infection in 2D segmentation, Tables \ref{seg_2D}, and abdomen1K~\cite{abdo_1K}, AMOS22~\cite{amos}, and ADCD~\cite{acdc} in 3D segmentation, Tables \ref{seg_3D}), as well as fine-grained and vascular-structure segmentation (OCT artery-vein~\cite{octa}, retinal vessels~\cite{hrf, drive, CHAOS}, and intercellular gaps~\cite{CREMI} in 2D segmentation, Tables \ref{seg_2D}, and pulmonary artery-vein~\cite{av} in 3D segmentation, Tables \ref{seg_3D}). We also provide the segmentation results in Fig.~\ref{seg} for a better-visualized comparison. In large-scale anatomical segmentations, Frepa shows a slight improvement over other models, including MedSAM, attributed to its enhanced ability to capture high-frequency details, thereby improving boundary identification. For the fine-grained segmentation tasks, Frepa exhibits more markedly superior performance, with a significant improvement ranging from +7$\%$ to +15$\%$ in DSC compared to the best-performing models. Unlike other models, which produce significantly less accurate and fragmented segmentations (Fig.\ref{moti}, Fig.\ref{seg}), Frepa achieves superior performance by capturing intricate vascular networks, preserving topological structures and fine-grained details, while the results maintain superior connectivity and closely align with the ground truth, while the results. Moreover, this advancement is preserved when applied to external modalities, such as the OCT reconstruction dataset and EM dataset, which exhibit entirely different imaging characteristics from other modalities. These findings robustly demonstrate Frepa's superior representation of fine-grained information and high-frequency components, along with better generalizability than other foundation models when applied to a broader range of realistic medical tasks.

\begin{table*}[!h]
\centering
\resizebox{\textwidth}{!}{
\arrayrulecolor{black}
\begin{tabular}{|c|c|c|c|c|c|c|c|c|}
\hline
\rowcolor[HTML]{EFEFEF} 
Modality                                                                       & Dataset                     & Metric     & MAE        & MFM         & MedSAM              & Frepa+ViT           & Frepa+SwinT         & Frepa+ConvNeXt      \\ \hline
                                                                               & Lung                & DSC($\%$) & 93.24$\pm$1.41 & 93.09$\pm$2.44  & 95.36$\pm$3.47          & \underline{96.89$\pm$1.19}    & 96.46$\pm$2.80          & \textbf{96.90$\pm$1.90} \\ \cline{2-9} 
                                                                               & Heart               & DSC($\%$) & 88.68$\pm$2.02 & 88.97$\pm$2.30   & \textbf{91.38$\pm$2.32} & \underline{91.30$\pm$2.70}    & 90.72$\pm$2.59          & 91.00$\pm$2.80          \\ \cline{2-9} 
\multirow{-3}{*}{CT}                                                           & Pneumonia infection & DSC($\%$) & 78.30$\pm$9.73  & 75.47$\pm$10.18 & 71.07$\pm$10.72         & \textbf{81.47$\pm$8.19} & \underline{81.51$\pm$9.35}    & 80.40$\pm$8.50          \\ \hline
                                                                               & OCTA500 artery      & DSC($\%$) & 74.86$\pm$5.94 & 64.35$\pm$7.29   & 71.59$\pm$5.13          & 78.00$\pm$6.44          & \underline{79.08$\pm$5.79}    & \textbf{79.80$\pm$5.90} \\ \cline{2-9} 
                                                                               & OCTA500 vein        & DSC($\%$) & 70.42$\pm$4.42 & 63.44$\pm$4.98  & 66.31$\pm$3.88          & 76.15$\pm$5.11          & \textbf{77.35$\pm$4.20} & \underline{77.00$\pm$4.10}    \\ \cline{2-9} 
                                                                               & OCT500 artery       & DSC($\%$) & 53.70$\pm$8.39 & 46.45$\pm$8.67  & 54.51$\pm$7.50          & 59.45$\pm$6.41          & \underline{61.88$\pm$6.95}    & \textbf{62.50$\pm$8.00} \\ \cline{2-9} 
\multirow{-4}{*}{\begin{tabular}[c]{@{}c@{}}OCT\\ reconstruction\end{tabular}} & OCT500 vein         & DSC($\%$) & 55.94$\pm$7.33 & 48.25$\pm$10.40 & 53.30$\pm$8.11          & 60.75$\pm$7.36          & \textbf{61.66$\pm$8.52} & \underline{60.90$\pm$7.00}    \\ \hline
                                                                               & HRF                 & DSC($\%$) & 62.02$\pm$4.16 & 54.21$\pm$4.22  & 72.39$\pm$2.33          & 79.88$\pm$2.27          & \underline{80.01$\pm$2.67}    & \textbf{80.50$\pm$2.50} \\ \cline{2-9} 
                                                                               & DRIVE               & DSC($\%$) & 78.75$\pm$3.30 & 70.82$\pm$4.03  & 80.52$\pm$4.13          & 82.90$\pm$3.84          & \underline{83.56$\pm$3.25}    & \textbf{84.00$\pm$3.10} \\ \cline{2-9} 
\multirow{-3}{*}{Retina}                                                       & CHASEDB             & DSC($\%$) & 70.96$\pm$5.53 & 68.67$\pm$2.19  & 74.84$\pm$1.91          & \underline{81.15$\pm$2.21}    & \textbf{81.52$\pm$2.06} & 80.80$\pm$2.50          \\ \hline
EM                                                                             & CREMI               & DSC($\%$) & 68.68$\pm$9.77 & 61.04$\pm$9.49  & 69.21$\pm$9.77          & \underline{78.70$\pm$9.40}    & \textbf{79.09$\pm$9.16} & 78.20$\pm$8.20          \\ \hline
\end{tabular}}
\caption{Performance on 2D image segmentation tasks. "DSC"=Dice similarity coefficient. OCT reconstruction and EM are two external modalities. \textbf{Bold} indicates the best results, and \underline{underline} indicates the second best results.}

\label{seg_2D}
\end{table*}

\begin{table*}[!h]
\centering
\resizebox{\textwidth}{!}{
\begin{tabular}{|c|c|c|c|c|c|c|c|c|}
\hline
\rowcolor[HTML]{EFEFEF} 
Modality              & Dataset          & Metric & MAE         & MFM          & MedSAM      & Frepa+ViT           & Frepa+SwinT          & Frepa+ConvNeXt      \\ \hline
                      & Pulmonary artery & DSC($\%$)    & 62.59$\pm$13.22 & 57.97$\pm$12.66  & 64.57$\pm$11.37 & 77.57$\pm$9.53          & \textbf{79.58$\pm$10.69} & \underline{79.08$\pm$10.15}   \\ \cline{2-9} 
                      & Pulmonary vein   & DSC($\%$)    & 57.57$\pm$10.37 & 52.44$\pm$11.73  & 61.71$\pm$11.70 & 75.65$\pm$10.99         & \underline{77.85$\pm$7.94}     & \textbf{77.95$\pm$8.12} \\ \cline{2-9} 
                      & Liver            & DSC($\%$)    & 94.00$\pm$2.02  & 91.66$\pm$1.64   & 94.88$\pm$1.38  & \textbf{95.14$\pm$1.37} & 94.54$\pm$1.96           & \underline{95.10$\pm$2.10}    \\ \cline{2-9} 
                      & Kidney           & DSC($\%$)    & 89.95$\pm$8.69 & 79.12$\pm$23.29 & 91.18$\pm$5.17  & 92.15$\pm$4.87          & \underline{92.47$\pm$3.87}     & \textbf{92.60$\pm$4.20} \\ \cline{2-9} 
\multirow{-5}{*}{CT}  & Spleen           & DSC($\%$)    & 91.80$\pm$5.61 & 89.07$\pm$7.08  & 94.90$\pm$1.50  & 95.36$\pm$1.11          & \underline{95.50$\pm$1.55}     & \textbf{95.55$\pm$1.80} \\ \hline
                      & AMOS22           & DSC($\%$)    & 71.23$\pm$5.13  & 70.60$\pm$6.27    & 73.98$\pm$3.18  & 76.39$\pm$4.34          & \underline{77.13$\pm$4.72}     & \textbf{77.50$\pm$4.90} \\ \cline{2-9} 
\multirow{-2}{*}{MRI} & ADCD             & DSC($\%$)    & 68.27$\pm$18.66 & 59.36$\pm$22.50  & 71.74$\pm$18.89 & 80.54$\pm$19.57         & \textbf{82.99$\pm$15.03} & \underline{82.50$\pm$14.80}   \\ \hline
\end{tabular}
}
\caption{Performance on 3D image segmentation tasks. "DSC"=Dice similarity coefficient. \textbf{Bold} indicates the best results, and \underline{underline} indicates the second best results.}
\label{seg_3D}
\end{table*}

\begin{table*}[!h]
\centering
\resizebox{0.8\textwidth}{!}{
\begin{tabular}{|c|c|c|c|c|c|c|}
\hline
\rowcolor[HTML]{EFEFEF} 
Modality              & Dataset          & Metric & 3D MAE      & 3DFrepa+ViT      & 3DFrepa+SwinT       & 3DFrepa+ConvNeXt    \\ \hline
                      & Pulmonary artery & DSC($\%$)    & 77.20$\pm$10.10 & 82.10$\pm$8.20       & \underline{84.80$\pm$9.50}    & \textbf{86.50$\pm$9.80} \\ \cline{2-7} 
                      & Pulmonary vein   & DSC($\%$)    & 74.30$\pm$9.01  & 79.10$\pm$2.80       & \textbf{85.50$\pm$7.00} & \underline{85.00$\pm$8.20}    \\ \cline{2-7} 
                      & Liver            & DSC($\%$)    & 95.50$\pm$1.80  & \underline{95.50$\pm$1.20} & 95.05$\pm$1.10          & \textbf{96.10$\pm$1.00} \\ \cline{2-7} 
                      & Kidney           & DSC($\%$)    & 91.80$\pm$2.50  & 92.70$\pm$3.79       & \underline{93.20$\pm$3.03}    & \textbf{93.80$\pm$3.28} \\ \cline{2-7} 
\multirow{-5}{*}{CT}  & Spleen           & DSC($\%$)    & 92.50$\pm$1.80  & 94.80$\pm$1.40       & \underline{96.00$\pm$1.66}    & \textbf{96.02$\pm$1.19} \\ \hline
                      & AMOS22           & DSC($\%$)    & 78.54$\pm$7.90  & 79.70$\pm$6.77        & \textbf{81.87$\pm$6.87} & \textit{80.50$\pm$5.33} \\ \cline{2-7} 
\multirow{-2}{*}{MRI} & ADCD             & DSC($\%$)    & 83.86$\pm$3.18  & 85.35$\pm$9.92       & \underline{84.08$\pm$7.18}    & \textbf{87.81$\pm$8.64} \\ \hline
\end{tabular}
}
\caption{Performance on 3D image segmentation tasks with 3D pretrained methods. "DSC"=Dice similarity coefficient. \textbf{Bold} indicates the best results, and \underline{underline} indicates the second best results.}
\label{3Dseg_3D}
\end{table*}

\subsection{Ablation study}
We perform the ablation study of the components in Frepa to evaluate their respective contributions. The validation is performed on datasets including Pneumonia MNIST~\cite{medmnist} for classification, Lung tumor~\cite{nodule_1} for lesion detection, and OCTA arteries and veins~\cite{octa} for segmentation. The seven most critical components of Frepa examined are high-frequency masking (HM), low-frequency perturbation (LP), frequency dual-component masking (FM), equalized-histogram image-domain masking (IM), HFL loss, embedding consistency loss (ConL), and gradient loss (GradL). For the ablation, these components are individually replaced or removed as follows: HM and LP are separately performed, FM branch is deleted, IM is replaced by zero-patch masking (Fig.~\ref{dist}, second column), HFL, ConL, and GradL are excluded during the pretraining phase. 

\begin{table*}[!h]
\centering
\resizebox{\textwidth}{!}{
\begin{tabular}{cccccccccc}
\hline
\rowcolor[HTML]{EFEFEF}
                &           & Frepa+ViT      & w/o HM & w/o LP & w/o FM & w/o IM & w/o HFL & w/o ConL & w/o GradL \\ \hline
                & ACC       & 0.934          & 0.931  & 0.927  & 0.921  & 0.931  & 0.927   & 0.933  & 0.933      \\
\multirow{-2}{*}{Pneumonia MNIST}                & AUC       & 0.970           & 0.968  & 0.966  & 0.965  & 0.972  & 0.967   & 0.970   & 0.968     \\ \hline
                & Detection & 0.951          & 0.943  & 0.938  & 0.916  & 0.919  & 0.939   & 0.945  & 0.949     \\
\multirow{-2}{*}{Lung tumor}                & IoU       & 0.656          & 0.634  & 0.623  & 0.590   & 0.607  & 0.624   & 0.632  & 0.655     \\ \hline
OCTA artery     & DSC($\%$)       & 78.00          & 73.61  & 76.48  & 73.48  & 76.31  & 74.69   & 77.70  & 74.95     \\
OCTA vein       & DSC($\%$)       & 76.15          & 73.04  & 73.55  & 71.71  & 73.89  & 73.11   & 76.03  & 74.29     \\ \hline
                &           &                &        &        &        &        &         &        &           \\

\hline
\rowcolor[HTML]{EFEFEF} 
                &           & Frepa+SwinT    & w/o HM & w/o LP & w/o FM & w/o IM & w/o HFL & w/o ConL & w/o GradL \\\hline
                & ACC       & 0.932          & 0.929  & 0.921  & 0.913  & 0.926  & 0.930    & 0.932  & 0.932      \\
\multirow{-2}{*}{Pneumonia MNIST}                & AUC       & 0.981          & 0.980   & 0.972  & 0.971  & 0.977  & 0.978   & 0.980   & 0.979     \\\hline
                & Detection & 0.948          & 0.942  & 0.935  & 0.905  & 0.919  & 0.928   & 0.933  & 0.941     \\
\multirow{-2}{*}{Lung tumor}                & IoU       & 0.677          & 0.658  & 0.642  & 0.604  & 0.621  & 0.624   & 0.655  & 0.676     \\\hline
OCTA artery     & DSC($\%$)       & 79.08          & 75.08  & 77.73  & 73.96  & 76.58  & 76.23   & 78.82  & 76.70      \\
OCTA vein       & DSC($\%$)       & 77.35          & 73.11  & 77.17  & 72.40  & 76.65  & 75.90   & 76.28  & 76.09     \\\hline
                &           &                &        &        &        &        &         &        &           \\
\hline                
\rowcolor[HTML]{EFEFEF} 
                &           & Frepa+ConvNeXt & w/o HM & w/o LP & w/o FM & w/o IM & w/o HFL & w/o ConL & w/o GradL \\\hline
                & ACC       & 0.934          & 0.931  & 0.927  & 0.921  & 0.931  & 0.927   & 0.933  & 0.933      \\
\multirow{-2}{*}{Pneumonia MNIS}                & AUC       & 0.970           & 0.968  & 0.966  & 0.965  & 0.972  & 0.967   & 0.970   & 0.970     \\\hline
                & Detection & 0.939          & 0.933  & 0.925  & 0.918  & 0.892  & 0.905   & 0.914  & 0.927     \\
\multirow{-2}{*}{Lung tumor}                & IoU       & 0.657          & 0.651  & 0.629  & 0.625  & 0.593  & 0.599   & 0.608  & 0.653     \\\hline
OCTA artery     & DSC($\%$)       & 79.80           & 76.19  & 78.45  & 75.13  & 78.12  & 76.75   & 78.53  & 76.97     \\
OCTA vein       & DSC($\%$)       & 77.00           & 72.91  & 75.39  & 72.50  & 75.48  & 74.00   & 76.11  & 75.75     \\ \hline
\end{tabular}}
\caption{The ablation study of Frepa. Our default setting is shown in gray color. "w/o"=without.}

\label{ablation}
\end{table*}

The results of the ablation study are presented in Table~\ref{ablation}. Removing each component leads to a decline in model performance across all tasks. Notably, the FM has the most significant impact; its removal causes the performance to degrade to be comparable to that of the MAE, suggesting a failure in high-frequency representation. Within the FM strategy, high-frequency masking and low-frequency perturbation serve distinct functions: for tasks that primarily depend on frequency features, such as Pneumonia MNIST classification, low-frequency perturbation plays a more critical role, whereas for tasks requiring fine-grained details, such as OCTA vessel segmentation, high-frequency masking becomes more essential. The results also indicate that the zero-patch masking is less effective than our proposed equalized-histogram IM strategy. Additionally, the exclusion of HFL loss, consistency loss, and gradient loss also contributes to a general reduction in performance. These findings are consistent for ViT, SwinT, and ConvNeXt architectures. In summary, the masking strategy and loss function designs are crucial components that significantly enhance the performance of Frepa.

\subsection{Frequency representation in image embedding}
\begin{figure}[!h]
\centering
\centerline{\includegraphics[width=0.95\columnwidth]{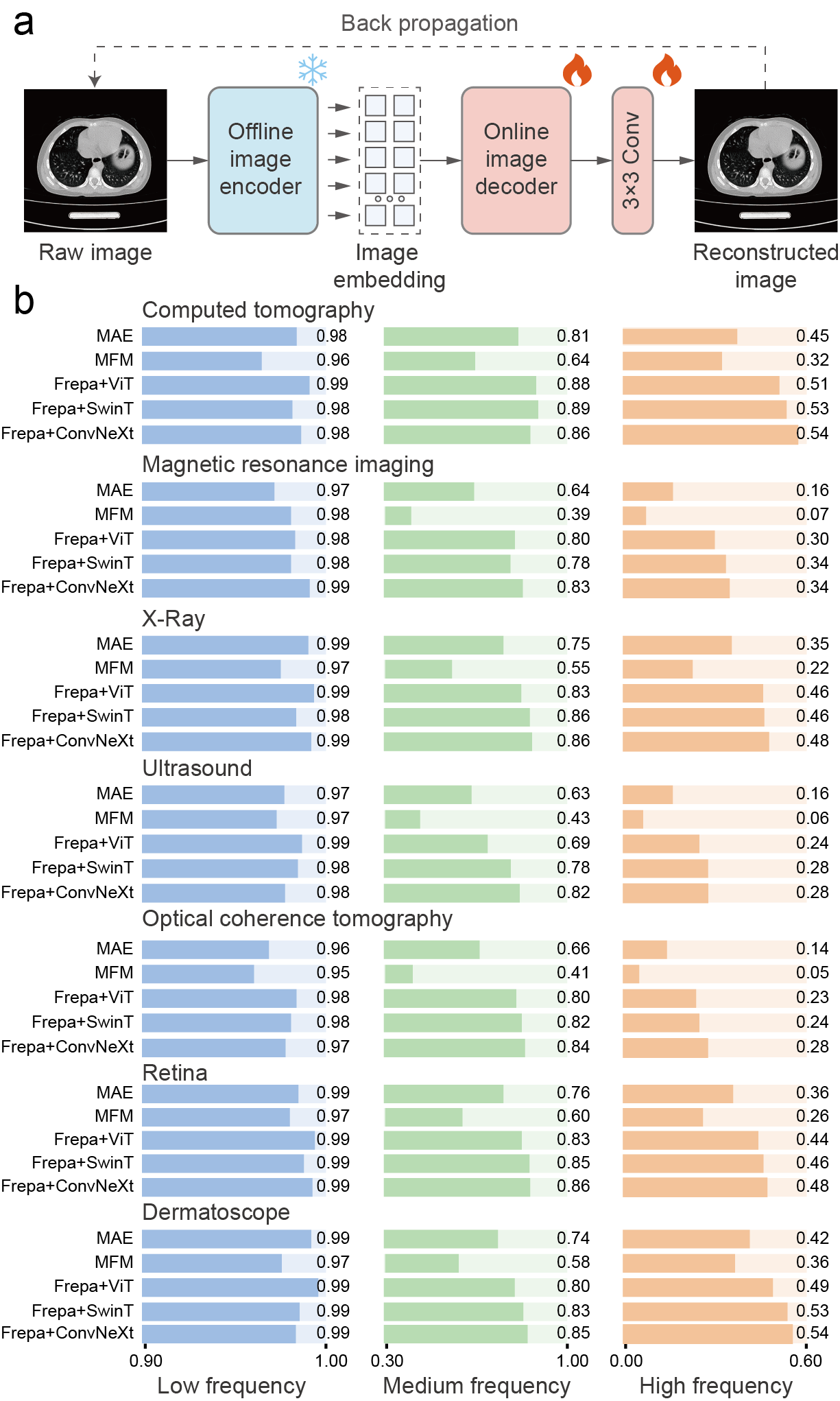}}
\caption{Comparison of different frequency component embeddings with various autoencoder pretraining approaches. (a) The pipeline of our designed experiment. (b) Quantitative comparison of the preserved components in embeddings for low (blue), medium (green), and high (orange) frequencies.}
\label{frequency}
\end{figure}

In this section, instead of using downstream tasks, we design a more straightforward approach to evaluate the representation and preservation of high-frequency components in the image embeddings. The experimental pipeline is illustrated in Fig.~\ref{frequency}a. In this setup, raw images are fed into fixed encoders (offline), and then decoders (online) are trained to reconstruct the raw images
\begin{equation}
    x^\prime=D_{on}\circ E_{off}\left( x \right) \rightarrow x.
\end{equation}
MSE loss is employed during training. Subsequently, we compare the reconstructed images with the raw images across different frequency components, utilizing 1-RMSE as the metric to measure the reconstruction similarity.

\begin{equation}
\rho _k= 1- \sqrt{\frac{1}{H\times W}\left\| O_k\left( x^\prime \right) -O_k\left( x \right) \right\| ^2},
\end{equation}
where $O_k \in \{O_l, O_m, O_h\}$ represents low-pass, medium-pass, and high-pass filtering, respectively. A higher value of $\rho _k$ indicates better restoration. We compared Frepa, implemented with ViT, SwinT, and ConvNeXt, against two other autoencoder methods: MAE and MFM. The results of this comparison are presented in Fig.~\ref{frequency}b. It is evident that Frepa consistently outperforms the other autoencoder methods in medium-frequency and high-frequency components while maintaining comparable performance in low-frequency components. This underscores Frepa’s superior ability to represent and preserve high-frequency information without compromising low-frequency representation. Among the three backbones, the performance of Swin Transformer and ConvNeXt shows slightly better preservation compared to ViT under medium and high-pass filters. This aligns with the observation that the pure ViT architecture tends to exhibit weaker representation capabilities in high-frequency domains. These findings align with our experiments on downstream tasks, providing a direct and intuitive evaluation of the level of embedding spaces.

\section{Discussion and Conclusion}
 Foundation models have achieved notable success with natural images by allowing the encoder to understand images from a global-level perspective. This approach is beneficial for tasks such as classification and object detection. However, in medical image analysis, this method faces significant challenges due to the anatomical similarities and low intra-modality variance among individuals~\cite{anno}, which result in less significance of global-level perception within one modality. Additionally, accurate diagnosis commonly depends on detailed information from specific anatomical structures, requiring more refined representations. Consequently, capturing only global-level and low-frequency information is inadequate for medical image analysis.

Our proposed Frepa addresses these limitations by enhancing high-frequency components and fine-grained features without sacrificing image-level perception. Our experiments demonstrate that Frepa can achieve a general performance improvement, especially for those requiring high-frequency details, surpassing both SOTA models and task-specific models. Our experiments further directly demonstrate that Frepa better captures and preserves high-frequency components from various perspectives, suggesting its potential use in other fields requiring fine-grained information, such as geospatial learning and spectral imaging. While our study shows excellent performance, future work could focus on optimizing Frepa for coarse-grained tasks and enhancing robustness to image noise, particularly in low-dose CT scans. Frepa's versatile and effective approach paves the way for advancements in medical AI foundation models.

\end{document}